\documentclass[twocolumn,showpacs,aps,prd,superscriptaddress,floatfix]{revtex4}
\usepackage{epsfig}
\usepackage{rotating}

\RequirePackage{xspace}

\newcommand{\DBR}{\mbox {$\Delta \ensuremath{\cal B}$}}
\def\smaller{\footnotesize}
\def\babar{\mbox{\slshape B\kern-0.1em{\smaller A}\kern-0.1em
    B\kern-0.1em{\smaller A\kern-0.2em R}}}
\def\invfb   {\ensuremath{\mbox{\,fb}^{-1}}\xspace}
\def\Y#1S{\ensuremath{\Upsilon{(#1S)}}\xspace}
\def\FourS {\Y4S}
\def\jpsi     {\ensuremath{{J\mskip -3mu/\mskip -2mu\psi\mskip 2mu}}\xspace}
\def\pep2{PEP-II}
\def\mev {\ensuremath{\mathrm{\,Me\kern -0.1em V}}\xspace}
\def\gev {\ensuremath{\mathrm{\,Ge\kern -0.1em V}}\xspace}
\def\gevcc {\ensuremath{{\mathrm{\,Ge\kern -0.1em V\!/}c^2}}\xspace}
\def\mes        {\mbox{$m_{ES}$}\xspace}
\def\DeltaE     {\mbox{$\Delta E$}\xspace}
\def\KL    {\ensuremath{K^0_{\scriptscriptstyle L}}\xspace} 
\def\BR {\ensuremath{\cal B}\xspace}
\def\rad{\ensuremath{\rm \,rad}\xspace}
\def\ulnu{$b \rightarrow u \ell \nu$}
\def\clnu{$b \rightarrow c \ell \nu$}

\def\pimunu{$B^{0} \rightarrow \pi^{-} \mu^{+} \nu$}
\def\pilnu{$B^{0} \rightarrow \pi^{-} \ell^{+} \nu$}
\def\etalnu{$B^{+} \rightarrow \eta \ell^{+} \nu$}
\def\etaplnu{$B^{+} \rightarrow \eta^{\prime} \ell^{+} \nu$}
\def\omegalnu{$B \rightarrow \omega \ell \nu$}
\def\etaetaplnu{$B^{+} \rightarrow \eta^{(\prime)} \ell^{+} \nu$}

\def\rholnu{$B \rightarrow \rho \ell \nu$}
\def\dlnu{$B \rightarrow D \ell \nu$}
\def\dstrlnu{$B \rightarrow D^* \ell \nu$}
\def\bfpilnu{${\BR}(B^{0} \rightarrow \pi^{-} \ell^{+} \nu)$}
\def\bfpiZlnu{${\BR}(B^{+} \rightarrow \pi^{0} \ell^{+} \nu)$}
\def\bfetalnu{${\BR}(B^{+} \rightarrow \eta \ell^{+} \nu)$}
\def\bfetaplnu{${\BR}(B^{+} \rightarrow \eta^{\prime} \ell^{+} \nu)$}
\def\Bz      {\ensuremath{B^0}\xspace}
\def\Bu      {\ensuremath{B^+}\xspace}
\def\Bbar    {\kern 0.18em\overline{\kern -0.18em B}{}\xspace}
\def\Bzb     {\ensuremath{\Bbar^0}\xspace}
\def\Bub     {\ensuremath{B^-}\xspace}
\def\BzBzb   {\ensuremath{\Bz {\kern -0.16em \Bzb}}\xspace}
\def\BpBm    {\ensuremath{\Bu {\kern -0.16em \Bub}}\xspace}
\def\to                 {\ensuremath{\rightarrow}\xspace}
\def\upsbzbz {\ensuremath{\FourS \to \BzBzb}\xspace}
\def\bfupsbzbz{${\BR}(\upsbzbz)$}
\def\upsbpbm {\ensuremath{\FourS \to \BpBm}\xspace}
\def\bfupsbpbm{${\BR}(\upsbpbm)$}
\def\bfrhoClnu{${\BR}(B^{0} \rightarrow \rho^{-} \ell^{+} \nu)$}
\def\bfrhoZlnu{${\BR}(B^{+} \rightarrow \rho^{0} \ell^{+} \nu)$}
\def\bfomegalnu{${\BR}(B^{+} \rightarrow \omega \ell^{+} \nu)$}
\def\bfDlnu{${\BR}(B \rightarrow D \ell \nu)$}
\def\bfDstrlnu{${\BR}(B \rightarrow D^* \ell \nu)$}
\def\bfDdblstrlnu{${\BR}(B \rightarrow D^{**} \ell \nu)$}
\def\bfetaetaplnu{${\BR}(B^{+} \rightarrow \eta^{(\prime)} \ell^{+} \nu)$}
\def\bfpilnuq{$\Delta {\BR}(B^{0} \rightarrow \pi^{-} \ell^{+} \nu,q^2)$}
\def\bfetalnuq{$\Delta {\BR}(B^{+} \rightarrow \eta \ell^{+} \nu,q^2)$}

\def\bfpilnuqq{$\DBR(q^2)$}

\def\lnrt{loose neutrino reconstruction technique}

\def\qq{$q^2$}
\def\qqr{$q^2$} 
\def\fplus{$f_+(q^2)$}
\def\vub{$|V_{ub}|$}
\def\nQ{12}

\def\udsctau{$u\overline{u}/d\overline{d}/s\overline{s}/c\overline{c}/\tau^+
\tau^-$}

\def\obb{other $B\overline{B}$}

\def\DeltaEDef{\ensuremath{\Delta E = (P_B \cdot P_{beams} - s/2) / \sqrt{s}}\xspace}
\def\FitRegion{\ensuremath{|\Delta E| < 1.0~\gev \mbox{ and } m_{ES} > 5.19~\gev}\xspace}
\newcommand{\gevsq}{\ensuremath{\mathrm{\,Ge\kern -0.1em V^2\!}}\xspace}

\def\BFval{$\left(1.42 \pm 0.05_{stat} \pm 0.07_{syst} \right) \times 10^{-4}$}
\def\BFetaval{$\left(0.36 \pm 0.05_{stat} \pm 0.04_{syst} \right) \times 10^{-4}$}
\def\BFetapval{$\left(0.24 \pm 0.08_{stat} \pm 0.03_{syst} \right) \times 10^{-4}$}

\def\VubFpzVal{$\left(8.5 \pm 0.3_{stat} \pm 0.2_{syst} \right) \times {10^{-4}}$}

\newcommand{\BADVERSION}{30}
\newcommand{\BABARPubYear}{10}
\newcommand{\BABARPubNumber}{025}
\newcommand{\SLACPubNumber}{14267}

\begin{document}

\preprint{\babar-PUB-\BABARPubYear/\BABARPubNumber} 
\preprint{SLAC-PUB-\SLACPubNumber} 

\begin{flushleft}
\babar\ Analysis Document \# 2080 Version \BADVERSION\\
 \babar-PUB-\BABARPubYear/\BABARPubNumber\\
 SLAC-PUB-\SLACPubNumber\\
 Phys. Rev. D{\bf 83}, 052011 (2011)\\
\end{flushleft}

\title{\large\bf\boldmath
Measurement of the \pilnu\ and \etaetaplnu\ Branching Fractions, the \pilnu\ 
and \etalnu\ Form-Factor Shapes, and Determination of \vub}

\author{P.~del~Amo~Sanchez}
\author{J.~P.~Lees}
\author{V.~Poireau}
\author{E.~Prencipe}
\author{V.~Tisserand}
\affiliation{Laboratoire d'Annecy-le-Vieux de Physique des Particules (LAPP), Universit\'e de Savoie, CNRS/IN2P3,  F-74941 Annecy-Le-Vieux, France}
\author{J.~Garra~Tico}
\author{E.~Grauges}
\affiliation{Universitat de Barcelona, Facultat de Fisica, Departament ECM, E-08028 Barcelona, Spain }
\author{M.~Martinelli$^{ab}$}
\author{D.~A.~Milanes}
\author{A.~Palano$^{ab}$ }
\author{M.~Pappagallo$^{ab}$ }
\affiliation{INFN Sezione di Bari$^{a}$; Dipartimento di Fisica, Universit\`a di Bari$^{b}$, I-70126 Bari, Italy }
% ...
\collaboration{The \babar\ Collaboration}
\noaffiliation

\date{\today}

\begin{abstract}
\clearpage
 We report the results of a study of the exclusive charmless semileptonic 
decays, $B^{+} \rightarrow \eta^{(\prime)} \ell^{+} \nu$ and $B^{0} \rightarrow
\pi^{-} \ell^{+} \nu$, undertaken with approximately 464 million 
$\ensuremath{B\kern 0.18em\overline{\kern -0.18em B}}$ pairs collected at the 
$\Upsilon(4S)$ resonance with the 
${\mbox{\slshape B\kern-0.1em{\smaller A}\kern-0.1em B\kern-0.1em{\smaller 
A\kern-0.2em R}}}$ detector. 
The analysis uses events in which the signal $B$ decays are reconstructed with 
a loose neutrino reconstruction technique. We obtain partial branching 
fractions for $B^{+} \rightarrow \eta \ell^{+} \nu$ and $B^{0} \rightarrow
\pi^{-} \ell^{+} \nu$ decays in three and 12 bins of $q^2$, respectively, 
from which we extract the $f_+(q^2)$ form-factor shapes and the total branching
fractions ${\ensuremath{\cal B}}(B^{+} \rightarrow \eta \ell^{+} \nu)$ 
$ = \left(0.36 \pm 0.05_{stat} \pm 0.04_{syst} \right) \times 10^{-4}$ and 
${\ensuremath{\cal B}}(B^{0} \rightarrow \pi^- \ell^{+} \nu)$
$ = \left(1.42 \pm 0.05_{stat} \pm 0.07_{syst} \right) \times 10^{-4}$. We also
measure ${\ensuremath{\cal B}}(B^{+} \rightarrow \eta^{\prime} \ell^{+} 
\nu)$ $ = \left(0.24 \pm 0.08_{stat} \pm 0.03_{syst} \right) \times 10^{-4}$.
We obtain values for the magnitude of the CKM matrix element 
$\ensuremath{|V_{ub}|}$ using three different QCD calculations. 
\end{abstract}


\maketitle  
\parskip=0.0cm
\abovecaptionskip=0.0cm
\belowcaptionskip=0.0cm

\section{Introduction}

 A precise measurement of the CKM matrix~\cite{CKM} element \vub\ will 
constrain the description of weak interactions and CP violation in the Standard
Model. The rate for exclusive charmless semileptonic decays involving a scalar 
meson is proportional to $|V_{ub}f_+(q^2)|^2$, where the form factor \fplus\ 
depends on \qq, the square of the momentum transferred to the lepton-neutrino 
pair. Values of \fplus\ are given by unquenched Lattice QCD (LQCD) 
calculations~\cite{HPQCD06,FNAL}, reliable only at large \qq\ ($\gtrsim$ 16 
\gevsq), and by Light Cone Sum Rules (LCSR) calculations~\cite{LCSR,singlet}, 
based on approximations only valid at small \qq\ ($\lesssim$ 16 \gevsq). The 
value of \vub\ can thus be determined by the measurement of partial branching 
fractions of charmless semileptonic $B$ decays. Extraction of the \fplus\ 
form-factor shapes from exclusive decays~\cite{PlusCC} such as 
\pilnu~\cite{PreviousVub} and \etaetaplnu\ may be used to test theoretical 
calculations~\cite{PDG10}. The values of the branching fractions (BF) of the 
\etaetaplnu\ decays will also improve our knowledge of the composition of 
charmless semileptonic decays and help constrain the size of the gluonic 
singlet contribution to the form factors for these decays~\cite{singlet}. 

 In this paper, we present measurements of the partial BFs \bfetalnuq\ and 
\bfpilnuq\ in three and \nQ\ bins of \qq, respectively, as well as the total 
BFs for all three decay modes. Values of the total BFs were previously reported
in Refs.~\cite{PreviousVub, CLEOpilnu2, BAD1380, BELLEPiRholnu, oldeta}. We use
the values of \bfpilnuqq\ for the \pilnu\ mode with form-factor 
calculations~\cite{HPQCD06, FNAL, LCSR} to obtain values of \vub. Values of 
\vub\ have previously been extracted from \pilnu\ measurements by 
CLEO~\cite{CLEOpilnu2}, \babar~\cite{PreviousVub, BAD1380, oldpi} and 
Belle~\cite{BELLEPiRholnu}. A very recent measurement by \babar~\cite{Jochen} 
will be discussed in Section VII
.

\section{Data Sample and Simulation}

 We use a sample of 464 million $B\overline{B}$ pairs corresponding to an 
integrated luminosity of 422.6~\invfb\ collected at the \FourS\ resonance with 
the \babar\ detector~\cite{ref:babar} at the \pep2\ asymmetric-energy $e^+e^-$ 
storage rings and a sample of 44~\invfb\ collected approximately 40 \mev\ below
the \FourS\ resonance (denoted ``off-resonance data''). Detailed Monte Carlo 
(MC) simulations are used to optimize the signal selections, to estimate the 
signal efficiencies, and to obtain the shapes of the signal and background 
distributions. MC samples are generated for $\Upsilon (4S) \rightarrow 
B\bar{B}$ events, $e^+e^- \rightarrow$ \udsctau\ (continuum) events, and 
dedicated $B\bar{B}$ samples containing \pilnu\ and \etaetaplnu\ signal decays.
The signal MC events are produced with the FLATQ2 generator~\cite{bad809} and 
are reweighted to reproduce the $f_+(q^2,\alpha, c_B)$ Becirevic-Kaidalov (BK) 
parametrization~\cite{BK}, where the values of the shape and normalization 
parameters, $\alpha$ and $c_B$, are taken from Ref.~\cite{PreviousVub}. The 
\babar\ detector's acceptance and response are simulated using the GEANT4 
package~\cite{ref:babar}.

\section{Event Reconstruction and Candidate Selection}

 We reconstruct the \pilnu\ and \etaetaplnu\ decays. The $\eta$ meson is 
reconstructed in the $\eta \rightarrow \gamma\gamma$ and 
$\eta \rightarrow \pi^+\pi^-\pi^0$ decay channels (combined BF of 62\%) while 
the $\eta^{\prime}$ is reconstructed in the $\eta^{\prime} 
\rightarrow \eta\pi^+\pi^-$ channel, followed by the $\eta \rightarrow 
\gamma\gamma$ decay (product BF of 17.5\%)~\cite{PDG08}. The $\eta^{\prime} 
\rightarrow \rho^0 \gamma$ decay channel suffers from large backgrounds and we 
do not consider it. We carry out an untagged analysis with a 
\lnrt~\cite{PreviousVub}, thereby obtaining a large candidate sample.

 Event reconstruction with the \babar\ detector is described in detail 
elsewhere~\cite{ref:babar}. Electrons (muons) are identified by their 
characteristic shower signatures in the electromagnetic calorimeter (muon 
detector), while charged hadrons are identified using the Cherenkov detector 
and $dE/dx$ measurements in the drift chamber. The average electron (muon) 
reconstruction efficiency is 93\% (70\%), while its misidentification 
probability is $< 0.2$\% ($< 1.5$\%). The neutrino four-momentum, 
$P_{\nu}=(|\vec{p}_{miss}|,\vec{p}_{miss})$, is inferred from the difference 
between the momentum of the colliding-beam particles $\vec{p}_{beams}$ and the 
vector sum of the momenta of all the particles detected in a single event 
$\vec{p}_{tot}$, such that $\vec{p}_{miss}=\vec{p}_{beams}-\vec{p}_{tot}$. To 
evaluate $E_{tot}$, the energy sum of all the particles, we assume zero mass 
for all neutrals since photons are difficult to disentangle from neutral 
hadrons and we take the mass given by the particle identification selectors for
the charged particles. In this analysis, we calculate the momentum transfer as 
$q^2=(P_{B}-P_{meson})^2$ instead of $q^2=(P_{\ell}+P_{\nu})^2$, where 
$P_B$, $P_{meson}$ and $P_{\ell}$ are the four-momenta of the $B$ meson, of the
$\pi$, $\eta$ or $\eta^{\prime}$ meson, and of the lepton, respectively. With 
this choice, the value of \qqr\ is unaffected by any misreconstruction of the 
rest of the event. Here $P_B$ has an effective value. To estimate this value, 
we first combine the lepton with a $\pi$, $\eta$ or $\eta^{\prime}$ meson to 
form the so-called $Y$ pseudoparticle. The angle, $\theta_{BY}$, between the 
$Y$ and $B$ momenta in the \FourS\ frame, can be determined by assuming 
$B \rightarrow Y\nu$. In this frame, the $Y$ momentum, the $B$ momentum and the
angle $\theta_{BY}$ define a cone with the $Y$ momentum as its axis and where 
the true $B$ momentum lies somewhere on the surface of the cone. The $B$ rest 
frame is thus known up to an azimuthal angle $\phi$ defined with respect to the
$Y$ momentum. The value of \qqr\ is then computed as the average of four \qqr\ 
values corresponding to four possible angles, $\phi$, $\phi+\pi/2$, $\phi+\pi$,
$\phi+3\pi/2$ rad, where the angle $\phi$ is chosen randomly and where the
four values of \qqr\ are weighted by the factor $\sin^2\theta_B$, $\theta_B$
being the angle between the $B$ direction and the beam direction in the \FourS\
frame~\cite{DstrFF}. We note that, $\theta_{BY}$ being a real angle, 
$|\cos\theta_{BY}| \leq 1$. We correct for the reconstruction effects on the 
\qq\ resolution (0.51 \gevsq) by applying an unregularized unfolding algorithm 
to the measured \qq\ spectra~\cite{Cowan}. 

\begin{figure}
\begin{center}
\epsfig{file=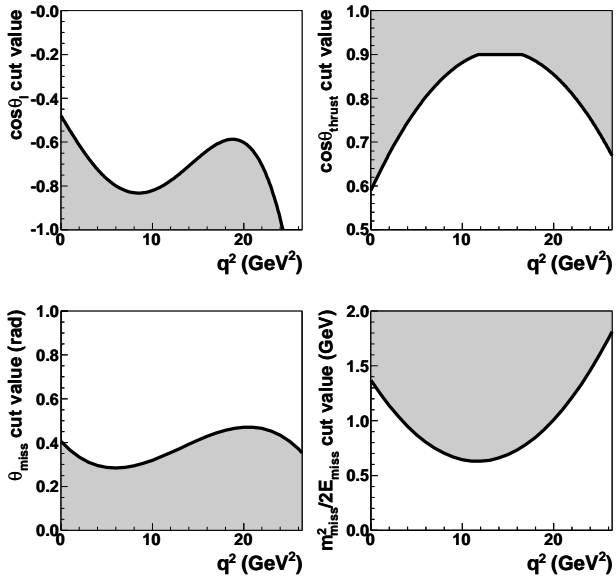,height=8.0cm}
\caption[]{\label{CUTJPG} 
Distributions of the selection values in the signal region for the 
$q^2$-dependent variables used in the analysis of \pilnu\ decays. 
The vertical axis represents the selection value for a given \qqr\ value. 
We reject an event when its value is in the shaded region.}
\end{center}
\end{figure}

\begin{figure}
\begin{center}
\epsfig{file=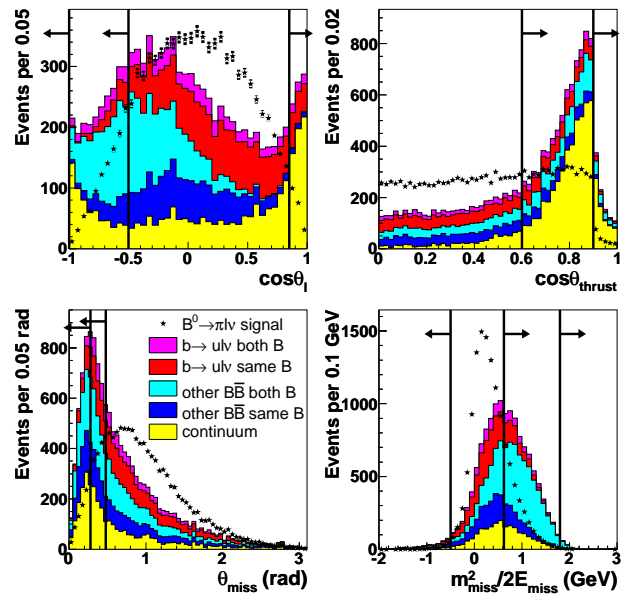,height=8.0cm}
\caption[]{\label{CutIllustration} (color online) Distributions in the signal
region for the $q^2$-dependent selections used in the analysis of \pilnu\ 
decays. The arrows indicate the rejected regions. All the selections have been 
applied except for the one of interest. In each panel, the signal area is 
scaled to the area of the total background.}
\end{center}
\end{figure}

 The candidate selections are optimized to maximize the ratio $S/\sqrt{(S+B)}$ 
in the MC simulation, where $S$ is the number of signal events and $B$ is the 
total number of background events. The continuum background is suppressed by 
requiring the ratio of second to zeroth Fox-Wolfram moments~\cite{FW} to be 
smaller than 0.5. This background is further suppressed for \pilnu\ by 
selections on the number of charged particle tracks and neutral calorimeter 
clusters~\cite{BhabhaVeto} that reject radiative Bhabha and converted photon 
processes. We ensure that the momenta of the lepton and meson candidates are 
kinematically compatible with a real signal decay by requiring that a 
geometrical vertex fit of the tracks of the two particles gives a $\chi^{2}$ 
probability greater than 1\% and that their angles in the laboratory frame be 
between 0.41 and 2.46 rad with respect to the $e^-$-beam direction, the 
acceptance of the detector. To avoid $\jpsi \rightarrow \mu^{+} \mu^{-}$ 
decays, we reject \pimunu\ candidates if the $Y$ mass corresponds to the \jpsi\
mass. The electron (muon) tracks are required to have momenta greater than 0.5 
(1.0) \gev\ in the laboratory frame to reduce the number of misidentified 
leptons and secondary decays such as $D \rightarrow X\ell\nu$, \jpsi, $\tau$ 
and kaon decays. Furthermore, the momenta of the lepton and the meson are 
restricted to enhance signal over background. We require the following: for 
\pilnu\ decays, $|\vec{p}^{_*}_{\ell}|>2.2$ \gev\ or $|\vec{p}^{_*}_{\pi}|>1.3$
\gev\ or $|\vec{p}^{_*}_{\ell}|+|\vec{p}^{_*}_{\pi}|>2.8$ \gev; for \etalnu\ 
decays, $|\vec{p}^{_*}_{\ell}|>2.1$ \gev\ or $|\vec{p}^{_*}_{\eta}|>1.3$ \gev\ 
or $|\vec{p}^{_*}_{\ell}|+|\vec{p}^{_*}_{\eta}|>2.8$ \gev; and for \etaplnu\ 
decays, $|\vec{p}^{_*}_{\ell}|>2.0$ \gev\ or 
$|\vec{p}^{_*}_{\eta^{\prime}}|>1.65$ \gev\ or 
$0.69\times|\vec{p}^{_*}_{\ell}|+|\vec{p}^{_*}_{\eta^{\prime}}|>2.4$ \gev\ 
(all asterisked variables are in the center-of-mass frame). For the 
\etaetaplnu\ decays, we restrict the reconstructed masses of the 
$\eta^{\prime}$ and $\eta$ to lie in the intervals 
$0.92<m_{\eta^{\prime}}<0.98$ \gev\ and $0.51<m_{\eta}<0.57$ \gev. For these 
decays, we also reject events with \qq\ higher than 16 \gevsq, since the signal
is dominated by background in that range. Most backgrounds are reduced by 
\qqr-dependent selections on the angle ($\cos\theta_{thrust}$) between the 
thrust axes of the $Y$ and of the rest of the event, on the polar angle 
($\theta_{miss}$) associated with $\vec{p}_{miss}$, on the invariant missing 
mass squared ($m^2_{miss}=E^2_{miss}- |\vec{p}_{miss}|^2$) divided by twice the
missing energy ($E_{miss}= E_{beams} - E_{tot}$), and on the angle 
($\cos\theta_{\ell}$) between the direction of the $W$ boson ($\ell$ and $\nu$
combined) in the rest frame of the $B$ meson and the direction of the lepton in
the rest frame of the $W$ boson. The \qq\ selections are shown in 
Fig.~\ref{CUTJPG} and their effects illustrated in Fig.~\ref{CutIllustration} 
for \pilnu\ decays. In Fig.~\ref{CutIllustration}, a single vertical line 
indicates a fixed cut; a set of two vertical lines represent a \qq-dependent 
cut. The position of the two lines correspond to the minimum and maximum values
of the cut, as shown in Fig.~\ref{CUTJPG}. The functions describing the \qq\ 
dependence are given in Tables~\ref{cutSummaryPi}-\ref{cutSummary} of the 
Appendix for the three decays under study. For \etalnu\ decays, additional 
background is rejected by requiring that $|\cos\theta_{V}|<0.95$, where 
$\theta_{V}$ is the helicity angle of the $\eta$ meson~\cite{bad809}. 

 The kinematic variables \DeltaEDef and $\ensuremath{m_{ES} = 
\sqrt{(s/2+\vec{p}_B \cdot \vec{p}_{beams})^2/E_{beams}^2 - \vec{p}_B^{\,2}}}$ 
are used in a two-dimensional extended maximum-likelihood fit~\cite{Barlow} to 
separate signal from background. Here, $\sqrt{s}$ is the center-of-mass energy 
of the colliding particles and $P_B = P_{meson}+P_{\ell}+P_{\nu}$, in the 
laboratory frame. We only retain candidates with \FitRegion, thereby removing 
a region with large backgrounds from the fit. On average, fewer than $1.14$ 
candidates is observed per event. For events with multiple candidates, only the
candidate with the largest value of $\cos\theta_{\ell}$ is kept. The signal 
event reconstruction efficiency varies between 8.3\% and 14.6\% for \pilnu, and
1.4\% and 2.6\% for \etalnu\ decays ($\gamma\gamma$ channel), depending on \qq.
It is 0.6\% for both \etalnu\ ($\pi^+\pi^-\pi^0$ channel) and \etaplnu\ decays.

\section{Backgrounds and Signal Extraction}

 Backgrounds can be broadly grouped into three main categories: decays arising 
from \ulnu\ transitions (other than the signal), decays in \obb\ events 
(excluding \ulnu) and decays in continuum events. For the \pilnu\ mode only, in
which there are many events, each of the first two categories is further split 
into a background category where the pion and the lepton come from the decay of
the same $B$ (``same $B$'' category), and a background category where the pion 
and the lepton come from the decay of different $B$ mesons (``both $B$'' 
category). 

 Given the sufficient number of events for the $\pi\ell\nu$ decay mode, the 
data samples can be subdivided in $12$ bins of \qqr\ for the signal and two 
bins for each of the five background categories. Two bins are used for each 
background category since the background \qq\ spectra are not that well known 
and need to be adjusted in the fit when the number of events is sufficiently 
large to permit it. The \qq\ ranges of the background binning for the \pilnu\ 
decays are: [0,18,26.4] \gevsq\ for the \ulnu\ same $B$ category, [0,22,26.4] 
\gevsq\ for the \ulnu\ both $B$ category, [0,10,26.4] \gevsq\ for the \obb\ 
same $B$ category, [0,14,26.4] \gevsq\ for the \obb\ both $B$ category and 
[0,22,26.4] \gevsq\ for the continuum category. In each case, the \qq\ ranges 
of the two bins are chosen to contain a similar number of events. All the 
signal and background events are fitted simultaneously for a total of 22 
parameters. For the $\eta^{(\prime)}\ell\nu$ modes, a smaller number of events 
leads us to restrict the signal and each of the three background categories to 
a single \qq\ bin except for the signal in the $\eta\ell\nu$ mode when 
$\eta \rightarrow \gamma\gamma$, which is investigated in three bins of \qqr. 

\begin{figure}
\begin{center}
\epsfig{file=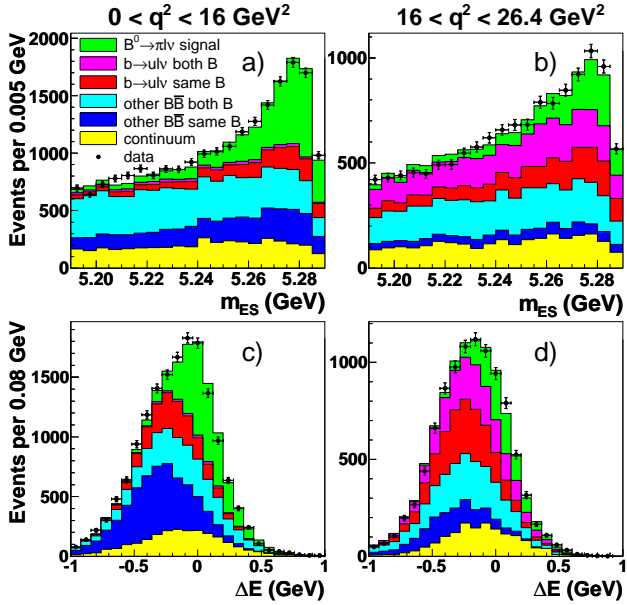,height=8.0cm}
\caption[]{\label{dEmESProjPilnu} (color online) Projections of the data and 
fit results for the \pilnu\ decays, in the signal-enhanced region: (a,b) \mes\ 
with $-0.16 < \Delta E < 0.20$ \gev; and (c,d) \DeltaE\ with \mes\ $>$ 5.268 
\gev. The distributions (a,c) and (b,d) are projections for \qqr\ $<$ 16 
\gevsq\  and for \qqr\ $>$ 16 \gevsq, respectively.}
\end{center}
\end{figure}

\begin{figure}
\begin{center}
\epsfig{file=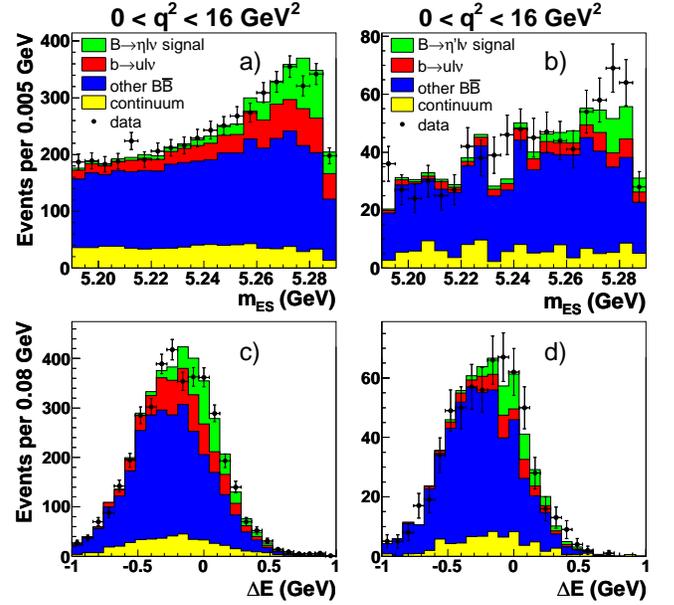,height=8.0cm, width=8.573cm}
\caption[]{\label{dEmESProjEta} (color online) Projections of the data and fit 
results for the \etaetaplnu\ decays, in the signal-enhanced region: (a,b) \mes\
with $-0.16 < \Delta E < 0.20$ \gev; and (c,d) \DeltaE\ with \mes\ $>$ 5.268 
\gev. The distributions (a,c) and (b,d) are projections for the \etalnu\ and 
\etaplnu\ decays, respectively, both for \qqr\ $<$ 16 \gevsq.}
\end{center}
\end{figure}

 We use the \DeltaE-\mes\ histograms, obtained from the MC simulation as 
two-dimensional probability density functions (PDFs), in our fit to the data 
to extract the yields of the signal and backgrounds as a function of \qqr. As 
an initial estimate, the MC continuum background yield and \qq-dependent shape 
are first normalized to match the yield and \qq-dependent shape of the 
off-resonance data control sample. This results in a large statistical 
uncertainty due to the small number of events in the off-peak data. To improve 
the statistical precision, the continuum background, initially normalized to 
the off-peak data, is allowed to vary in the fit to the data for the 
$\pi\ell\nu$ and $\eta\ell\nu (\gamma\gamma)$ modes where we have a large 
number of events. The fit result is compatible with the off-peak prediction 
within at most 1 standard deviation. Because of an insufficient number of 
events, the \ulnu\ background is fixed in the fit for the 
$\eta^{(\prime)}\ell\nu$ modes, and the continuum contribution is also fixed 
for the $\eta\ell\nu (3\pi)$ and $\eta^{\prime}\ell\nu$ modes. Whenever a 
background is not varied in the fit, it is fixed to the MC prediction except 
for the continuum background which is fixed to its normalized yield and 
\qq-dependent shape using the off-resonanc data. The background parameters 
which are free in the fit require an adjustment of less than 10\% with respect 
to the MC predictions. For illustration purposes only, we show in 
Fig.~\ref{dEmESProjPilnu} \DeltaE\ and \mes\ fit projections in the 
signal-enhanced region for \pilnu\ decays in two ranges of \qq\ corresponding 
to the sum of eight bins below and four bins above
\qqr\ = 16 \gevsq, respectively. More detailed \DeltaE\ and \mes\ fit 
projections in each \qqr\ bin are also shown in Figs.~\ref{dEFitProjDataPi} 
and~\ref{mESFitProjDataPi} of the Appendix for the \pilnu\ decays. The data and
the fit results are in good agreement. Fit projections for \etaetaplnu, only 
available below \qqr\ = 16 \gevsq, are shown in Fig.~\ref{dEmESProjEta}. Table 
\ref{yieldBGtable} gives the total fitted yields in the full \qq\ range for the
signal and each background category as well as the $\chi^2$ values and degrees 
of freedom for the overall fit region. The yield values in the \etalnu\ column 
are the result of the fit to the combined $\gamma\gamma$ and $3\pi$ modes. 

\begin{table}
\caption[]{\label{yieldBGtable} Fitted yields in the full \qq\ range for the 
signal and each background category, total number of MC and data events, and 
values of $\chi^2$ for the overall fit region.} 
\begin{center}
\begin{tabular}{lp{0.1cm}cp{0.1cm}cp{0.1cm}c}
\hline\hline
Decay mode          & &$\pi^-\ell^+\nu$& &$\eta\ell^+\nu$& &$\eta^{\prime}
\ell^+\nu$ \\ \hline
Signal              & & $11778\pm 435$ & &$888\pm98$    & &$141\pm46$ \\
\ulnu\              & & $27793\pm 929$ & &$2201 (fixed)$& &$204 (fixed)$ \\   
Other $B\overline{B}$& & $80185\pm 963$ & &$17429\pm247$ & &$2660\pm82$ \\ 
Continuum           & & $27790\pm 814$ & &$3435\pm195$  & &$517 (fixed)$ \\
\hline
MC events           & & $147546\pm467$ & &$23953\pm183$ & &$3522\pm68$ \\
Data events         & & $147529\pm384$ & &$23952\pm155$ & &$3517\pm59$  \\
$\chi^2$/ndf        & & $411/386$     & &$56/52$        & &$19/17$ \\
\hline\hline
\end{tabular}
\end{center}
\end{table}

\section{Systematic Uncertainties}

 Systematic uncertainties on the values of the partial branching fractions, 
\bfpilnuqq, and their correlations among the \qq\ bins have been investigated. 
These uncertainties are estimated from the variations of the resulting partial 
BF values (or total BF values for \etaplnu\ decays) when the data are 
reanalyzed with different simulation parameters and reweightings. For each 
parameter, we use the full MC dataset to generate new \DeltaE-\mes\ 
distributions (``MC event samples'') by varying randomly only the parameter of 
interest over a complete ($> 3 \sigma$) Gaussian distribution whose standard 
deviation is given by the uncertainty on the specific parameter under 
investigation. One hundred such samples are generated for each parameter. 
Uncertainties due to $B$ counting and final state radiation are estimated by 
generating only one sample. Each MC sample is analyzed the same way as real 
data to determine values of \bfpilnuqq\ (or total BF values for \etaplnu\ 
decays). The contribution of the parameter to the systematic uncertainty is 
given by the RMS value of the distribution of these values over the one hundred
samples. 

\begin{table*}
\caption[]{\label{errorsPi} Values of signal yields, \bfpilnuqq\ and their 
relative uncertainties (\%) for \pilnu, \etalnu\ and \etaplnu\ decays.}
\begin{center}
\begin{tabular}{lp{0.1cm}cp{0.1cm}cp{0.1cm}cp{0.1cm}cp{1cm}cp{1cm}c} 
\hline\hline Decay mode & & \multicolumn{7}{c}{$\pi^-\ell^+\nu$} & & $\eta\ell^+\nu$ & & $\eta^{\prime}\ell^+\nu$ \\ 
\hline \qq\ range (\gevsq) & &\qq$<$12  & &\qq$<$16 & &\qq$>$16 & & full \qq\ range & & \qq$<$16 & & \qq$<$16\\ 
\hline
Yield 		             &    & 6541.6 & &8422.1& & 3355.4& & 11777.6 & & 887.9 & & 141.0\\
BF ($10^{-4}$)               &    & 0.83  & & 1.09 & & 0.33  & & 1.42    & & 0.36  & & 0.24 \\
\hline
Statistical error                    &    & 3.9  & & 3.7 & & 7.6 & & 3.5        & & 12.5  & & 32.8 \\ 
Detector effects             &    & 3.1  & & 3.5 & & 6.1 & & 4.0        & & 8.0   & & 8.8 \\ 
Continuum bkg                &    & 0.9  & & 0.8 & & 1.0 & & 0.7        & & 0.3   & & 7.1  \\ 
$B \rightarrow X_u \ell \nu$ bkg& & 2.0  & & 1.7 & & 4.2 & & 2.0        & & 7.6   & & 6.7  \\ 
$B \rightarrow X_c \ell \nu$ bkg& & 0.6  & & 0.7 & & 1.8 & & 1.0        & & 1.2   & & 2.6  \\ 
Other effects                &    & 2.3  & & 2.2 & & 3.2 & & 2.3        & & 3.4   & & 4.6  \\ 
\hline
Total uncertainty            &    & 5.9  & & 5.9 & & 11.3 & & 6.3       & & 17.0  & & 35.8 \\ 
\hline\hline 
\end{tabular}
\end{center}
\end{table*}

  The systematic uncertainties due to the imperfect description of the detector
in the simulation are computed by using the uncertainties, determined from 
control samples, on the tracking efficiency of all charged particle tracks, 
on the particle identification efficiencies of signal candidate tracks, on the 
calorimeter efficiencies (varied separately for photons and \KL), on the energy
deposited in the calorimeter by \KL mesons as well as on their production 
spectrum. The reconstruction of these neutral particles affects the analysis 
through the neutrino reconstruction. The uncertainties due to the 
generator-level inputs to the simulation are given by the uncertainties in the 
BFs of the background processes \ulnu\ and \clnu, in the BFs of the secondary 
decays producing leptons~\cite{PDG08}, and in the BFs of the $\Upsilon(4S) 
\rightarrow B\overline{B}$ decays~\cite{PDG10}. The $B \rightarrow X \ell\nu$ 
form-factor uncertainties, where $X = (\pi,\rho,\omega,\eta^{(\prime)},D,D^*)$,
are given by recent calculations or measurements~\cite{PDG08}. The 
uncertainties in the heavy quark parameters used in the simulation of 
nonresonant \ulnu\ events are given in Ref.~\cite{Henning}. We assign an 
uncertainty of 20\%~\cite{photosErr} to the final state radiation (FSR) 
corrections calculated by PHOTOS~\cite{photos}. Finally, the uncertainties due 
to the modeling of the continuum are established by using the uncertainty in 
its \qqr\ distribution shape and, when the continuum background is fixed, the 
uncertainty in the total yield, both given by comparisons with the 
off-resonance data control sample.

  The list of all the systematic uncertainties, as well as their values for the
partial and total BFs, are given in Tables~\ref{pierror} and~\ref{etaerror} of 
the Appendix. The item ``Signal MC stat error'' in these tables includes the 
systematic uncertainty due to the unfolding procedure. The correlation matrices
obtained in the measurement of the partial BFs are presented in 
Tables~\ref{StatCovMCEta},~\ref{StatCovMC} and~\ref{SystCovMC}. A condensed 
version of all the uncertainties is given in Table \ref{errorsPi} together with
signal yields and partial BFs in selected \qqr\ ranges. The values given for 
the \etalnu\ decays are those obtained from the fits to the distributions of 
the $\eta \rightarrow \gamma \gamma$ and $\eta \rightarrow \pi^+ \pi^- \pi^0$ 
channels combined. The larger relative uncertainties occurring in bin 12 of 
Table~\ref{pierror} are due to poorly reconstructed events, and to the small 
raw yield in that bin. The former arises from the presence of a large number of
low momentum pions and a large background. This makes it difficult to select 
the right pion and results in a larger absolute uncertainty on the fitted 
yield. The small yield leads to a fairly large unfolding correction in this bin
and thus to a considerably reduced unfolded yield. On the other hand, the 
unfolding process increases the absolute uncertainty only slightly. The reduced
yield together with the larger absolute uncertainty lead to the larger relative
uncertainties reported in the table.

\section{Results}

\begin{figure*}[t]
\begin{minipage}[t]{.48\linewidth}
\centering\includegraphics[width=\linewidth,height=8.1cm]{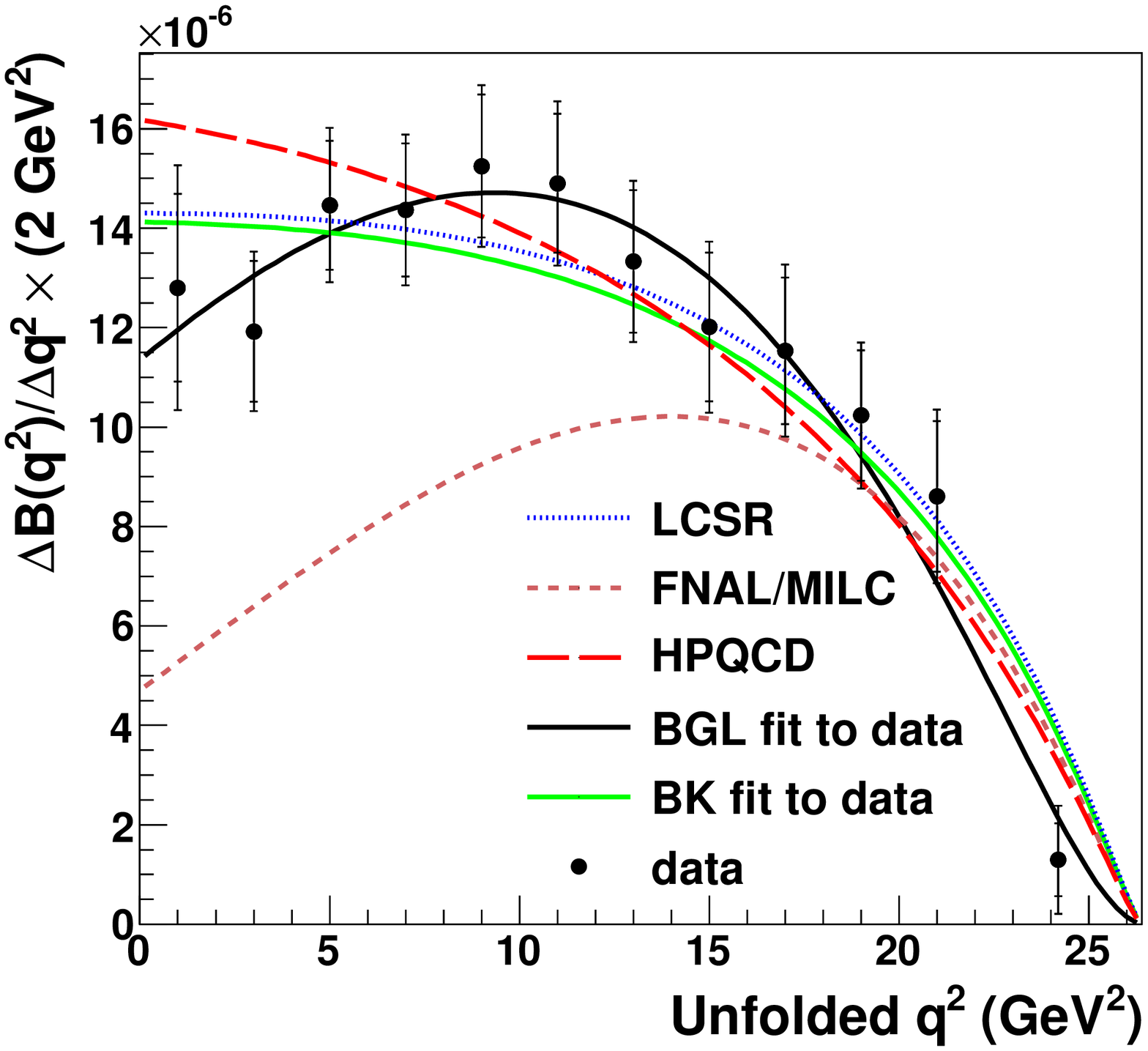}
\caption[]{\label{figFplus} 
(color online) Partial \bfpilnuqq\ spectrum in 12 bins of \qq\ for \pilnu\ 
decays. The data points are placed in the middle of each bin whose width is 
defined in Table~\ref{pierror}.The smaller error bars are statistical only 
while the larger ones also include systematic uncertainties. The solid green 
and black curves show the result of the fit to the data of the BK~\cite{BK} and
BGL~\cite{BGL} parametrizations, respectively. The data are also compared to 
unquenched LQCD calculations (HPQCD~\cite{HPQCD06}, FNAL~\cite{FNAL}) and a 
LCSR calculation~\cite{LCSR}.}
\end{minipage}\hfill
\begin{minipage}[t]{.48\linewidth}
\centering\includegraphics[width=\linewidth,height=8.1cm]{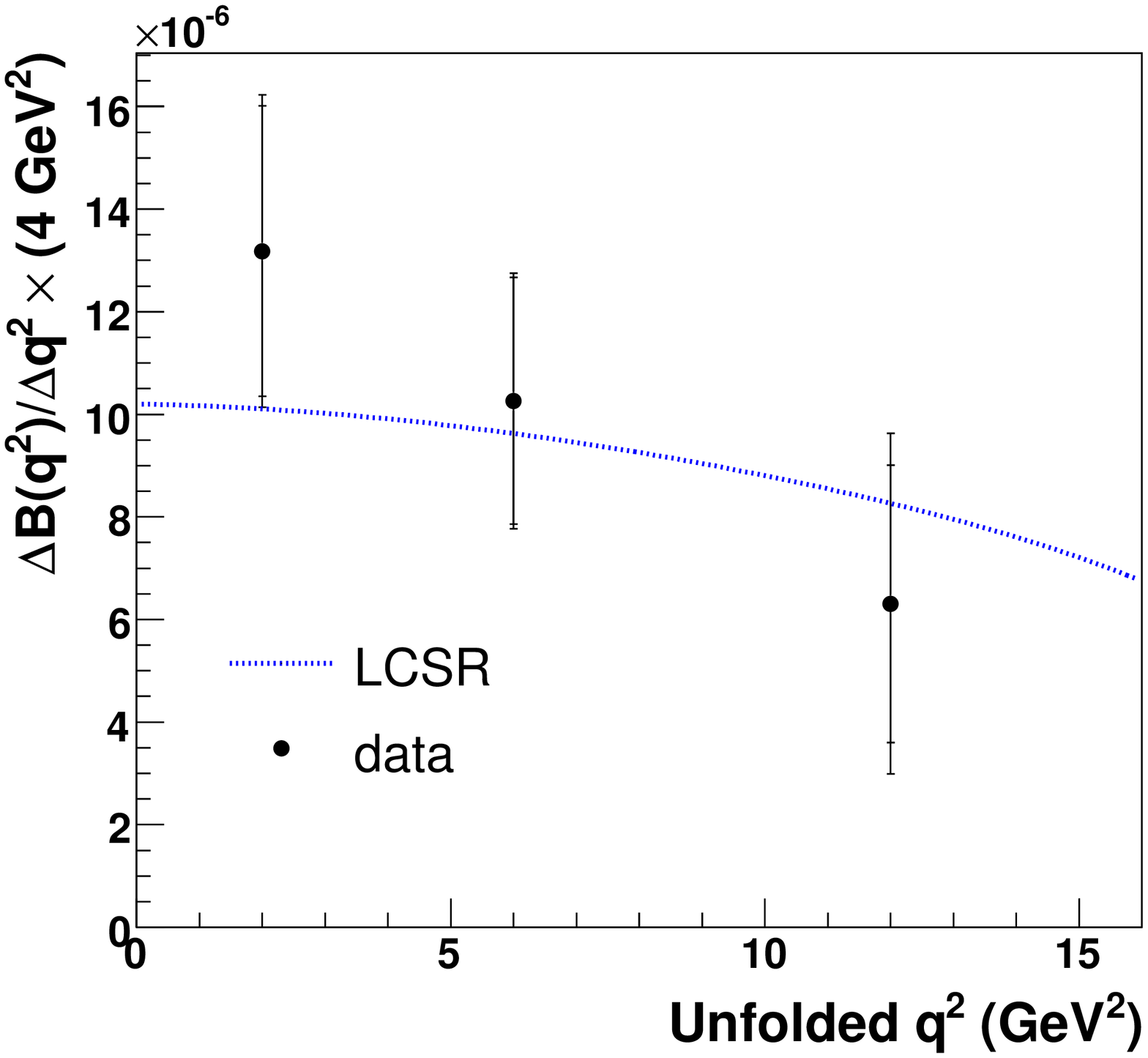}
\caption[]{\label{figFplusEta} 
(color online) Partial \bfpilnuqq\ spectrum in 3 bins of \qq\ for \etalnu\ 
decays. The data points are placed in the middle of each bin whose width is 
defined in Table~\ref{etaerror}. The smaller error bars are statistical only 
while the larger ones also include systematic uncertainties. The data are also 
compared to a LCSR calculation~\cite{singlet}.}  
\end{minipage}
\end{figure*}

 The partial BFs are calculated for \pilnu\ and \etalnu\ decays using the 
unfolded signal yields, the signal efficiencies given by the simulation and the
BFs \bfupsbzbz\ $= 0.484\pm0.006$ and 
\bfupsbpbm\ $= 0.516\pm0.006$~\cite{PDG10}. We obtain the total BFs \bfpilnu\ =
\BFval, \bfetalnu\ = \BFetaval\ and \bfetaplnu\ = \BFetapval. The BF value for 
\etaplnu\ has a significance of $3.2\sigma$ when we take into account only the
statistical uncertainty~\cite{signif}. Taking into account the effect of the 
systematic uncertainty which increases the total uncertainty by about 8\% 
leads to a reduced significance of $3.0\sigma$. The BF value, obtained from a 
fit to the combined $\gamma\gamma$ and $3\pi$ channels of the \etalnu\ decays, 
is in good agreement with the weighted average of the total BFs obtained 
separately for the $\gamma\gamma$ and $3\pi$ channels. Consistent results are 
obtained when dividing the final data set into chronologically-ordered subsets,
electron only and muon only subsets, modifying the \qq\ or the \DeltaE\ and 
\mes\ binnings, and varying the event selection requirements. 

The experimental \bfpilnuqq\ distributions are displayed in Fig.~\ref{figFplus}
for \pilnu\ decays and in Fig.~\ref{figFplusEta} for \etalnu\ decays, together 
with theoretical predictions. To allow a direct comparison with the theoretical
predictions, which do not include FSR effects, the experimental distributions 
in these figures have been obtained with the efficiency given by the ratio of
$q^2$ unfolded events generated after all the cuts with a simulation which 
includes FSR to the total number of events generated before any cut and with 
no FSR effects i.e. with PHOTOS switched off. We obtain the \fplus\ shape from
a fit to these distributions. The $\chi^2$ function minimized in the fit to the
\fplus\ shape uses the Boyd-Grinstein-Lebed (BGL) parametrization~\cite{BGL} 
consisting of a  two-parameter polynomial expansion. For the \pilnu\ decays, 
the fit gives $a_1/a_0 = -0.63 \pm 0.29$ and $a_2/a_0 = -6.9 \pm 1.7$, with 
$P(\chi^2)=92.1\%$ as well as a value of $|V_{ub}f_{+}(0)|$ = \VubFpzVal\ from 
the fit extrapolated to \qq\ $= 0$. This value can be used to predict rates of 
other decays such as $B \rightarrow \pi \pi$~\cite{f0}. For completeness, we 
also show the fit to the BK parametrization~\cite{BK}, which gives 
$\alpha_{BK}=0.52\pm0.04$, with $P(\chi^2)=28.6\%$.

\begin{table*}
\caption[]{\label{vubtable} Values of \vub\ derived from the form-factor 
calculations for the \pilnu\ decays. The three uncertainties on $|V_{ub}|$ are 
statistical, systematic and theoretical, respectively.}
\begin{center}
\begin{tabular}{ccccccc}
\hline\hline
         & $q^2$ (\gevsq) &$\DBR$ ($10^{-4}$) &$\Delta\zeta$ (ps$^{-1}$) & $|V_{ub}|$ ($10^{-3}$) & $\chi^2$/ndf & $Prob(\chi^2)$ \\ \hline
HPQCD~\cite{HPQCD06}      & $> 16$ & $0.33\pm 0.03\pm 0.03$ & $2.02\pm 0.55$ & $3.28\pm 0.13\pm 0.15{}^{+0.57}_{-0.37}$ & 5.0/4 & 28.8\%\\
FNAL~\cite{FNAL}    & $> 16$ & $0.33\pm 0.03\pm 0.03$ & $2.21{}^{+0.47}_{-0.42}$ & $3.14\pm 0.12\pm 0.14{}^{+0.35}_{-0.29}$ & 6.4/4 & 17.4\%\\
LCSR~\cite{LCSR} & $< 12$ & $0.84\pm 0.03\pm 0.04$ & $4.00{}^{+1.01}_{-0.95}$ & $3.70\pm 0.07\pm 0.08{}^{+0.54}_{-0.39}$ & 6.2/6 & 39.9\%\\

\hline\hline
\end{tabular}
\end{center}
\end{table*}

 The \qq\ distribution extracted from our data is compared in 
Fig.~\ref{figFplus} to the shape of the form factors obtained from the three
theoretical calculations listed in Table~\ref{vubtable}: the one based on Light
Cone Sum Rules~\cite{LCSR} for $q^2 < 12$ \gevsq, and the two based on 
unquenched LQCD~\cite{HPQCD06, FNAL} for $q^2 > 16$ \gevsq. We first normalize 
the form-factor predictions to the experimental data by requiring the
integrals of both to be the same over the \qqr\ ranges of validity given in 
Table \ref{vubtable} for each theoretical prediction. Considering only 
experimental uncertainties, we then calculate the $\chi^2$ probabilities 
relative to the binned data result for various theoretical predictions. These 
are given in Table~\ref{vubtable} for the \pilnu\ decays. All three 
calculations are compatible with the data. As shown in Fig. \ref{figFplusEta}, 
a LCSR calculation~\cite{singlet} is compatible with the data for the \etalnu\
decays. It should be noted that the theoretical curves in Fig.~\ref{figFplus} 
have been extrapolated over the full \qqr\ range based on a parametrization 
obtained over their \qqr\ ranges of validity. These extended ranges are only 
meant to illustrate a possible extension of the present theoretical 
calculations.

\begin{table*}
\caption[]{\label{average} Values of quantities of interest and their 
averages obtained in the study of \pilnu\ decays. The third uncertainty,
given for the average values, is due to the form-factor calculation. It is
not shown for the individual determination of $|V_{ub}|$. The results for
$a_1/a_0$, $a_2/a_0$ and $|V_{ub}f_+(0)|$ in the column titled ``Average'' 
are actually from a fit to the combined data, as discussed in the text.}
\begin{center}
\begin{tabular}{lp{0.1cm}cp{0.1cm}cp{0.1cm}c}
\hline\hline
&& Present work && Reference~\cite{Jochen} && Average \\ \hline
Total BF && $1.42 \pm 0.05 \pm 0.07$ && $1.41 \pm 0.05 \pm 0.07$ && $1.42 \pm 
0.04 \pm 0.07$ \\
$|V_{ub}|_{HPQCD}\times 10^{3}$ && $3.28 \pm 0.13 \pm 0.15$ && 
$3.21 \pm 0.13 \pm 0.12$ && $3.23 \pm 0.09 \pm 0.13{}^{+0.57}_{-0.37}$ \\
$|V_{ub}|_{FNAL}\times 10^{3}$ && $3.14\pm 0.12 \pm 0.14$ && 
$3.07 \pm 0.11 \pm 0.11$ && $3.09 \pm 0.08 \pm 0.12{}^{+0.35}_{-0.29}$ \\
$|V_{ub}|_{LCSR}\times 10^{3}$ && $3.70 \pm 0.07 \pm 0.08$ && 
$3.78 \pm 0.08 \pm 0.10$ && $3.72 \pm 0.05 \pm 0.09{}^{+0.54}_{-0.39}$ \\
$|V_{ub}f_+(0)|\times 10^{4}$ && $8.5 \pm 0.3 \pm 0.2$ && $10.8 \pm 0.5 \pm 
0.3$ && $9.4 \pm 0.3 \pm 0.3$ \\
BGL $a_1/a_0$ && $-0.63 \pm 0.27 \pm 0.10$ && $-0.82 \pm 0.23 \pm 0.17$ && 
$-0.79 \pm 0.14 \pm 0.14$ \\
BGL $a_2/a_0$ && $-6.9 \pm 1.3 \pm 1.1$    && $-1.1 \pm 1.6 \pm 0.9 $  &&
$ -4.4 \pm 0.8 \pm 0.9$ \\

\hline\hline
\end{tabular}
\end{center}
\end{table*}

 We extract a value of \vub\ from the \pilnu\ \bfpilnuqq\ distributions using 
the relation: $|V_{ub}| = \sqrt{\Delta{\cal B}/(\tau_{B^0}\Delta \zeta)}$, 
where $\tau_{B^0} = 1.525 \pm 0.009$ ps~\cite{PDG10} is the $B^0$ lifetime and 
$\Delta\zeta = \Gamma/|V_{ub}|^2$ is the normalized partial decay rate 
predicted by the form-factor calculations~\cite{HPQCD06, FNAL, LCSR}. The
quantities $\Delta{\cal B}$ and $\Delta \zeta$ are restricted to the \qqr\ 
ranges of validity given in Table \ref{vubtable}. The values of $\Delta \zeta$ 
are independent of experimental data. The values of \vub\ given in 
Table~\ref{vubtable} range from $(3.1-3.7)\times 10^{-3}$. A value of \vub\ 
could not be obtained from the \etalnu\ decays because the required theoretical
input $\Delta \zeta$ is not yet available.

\section{Combined \babar\ results}

 At first glance, there appears to be a large overlap between the present 
analysis of the \pilnu\ data and that of another recent \babar\ 
measurement~\cite{Jochen}. However, there are significant differences between 
the two analyses. Considering the same fit region, we obtain 147529 selected
events (signal or background) compared to 42516 such events in 
Ref.~\cite{Jochen}. This difference can easily be explained by the fact that we
use the full \babar\ data set in the present analysis but this is not so in 
Ref.~\cite{Jochen}. Furthermore, the use of the loose neutrino reconstruction 
technique in this work leads to a larger background. Only 140 events are found 
in common between the two data sets, i.e. 0.3\% overlap. The statistical 
uncertainties are thus expected to be uncorrelated between the two analyses. 
The event reconstruction and simulation are also somewhat different. For 
example, the values of $q^2$ are computed using different, although in 
principle equivalent, relations: here, $q^2 = (P_B - P_{\pi})^2$ versus 
$q^2 = (P_{\ell} + P_{\nu})^2$ in Ref.~\cite{Jochen}. Nevertheless, almost all 
of the systematic uncertainties are expected to be highly correlated.

 It is gratifying to note that, as shown in Table~\ref{average}, the total BF 
as well as the values of $|V_{ub}|$ obtained in the two analyses are in good 
agreement with each other. The value of \vub\ quoted under Ref.~\cite{Jochen}
in Table~\ref{average} for the FNAL~\cite{FNAL} theoretical prediction is 
obtained using the values of the partial BFs given in Ref.~\cite{Jochen} for
\qq\ $ > 16$ \gevsq. Similar numbers of signal events ($11778 \pm 435$ here
compared with $10604 \pm 376$ in Ref.~\cite{Jochen} when the events from 
$B^+ \rightarrow \pi^0\ell^+\nu$ decays are also included) lead to similar 
statistical uncertainties in the two analyses.

 It is possible to obtain a good approximation to the average of the present 
results and those of Ref.~\cite{Jochen} obtained in the \pilnu\ decays by 
taking the statistical uncertainties to be uncorrelated and the systematic 
uncertainties to be fully correlated. Such an averaging procedure yields the 
averages, and associated uncertainties, given in Table~\ref{average} for the 
total branching fraction and the values of $|V_{ub}|$. The additional value of
$|V_{ub}|$ obtained in Ref.~\cite{Jochen} with a combined fit to data and 
theoretical points is not taken into account in this computation.

\begin{figure}[!htb]
\begin{center}
\epsfig{file=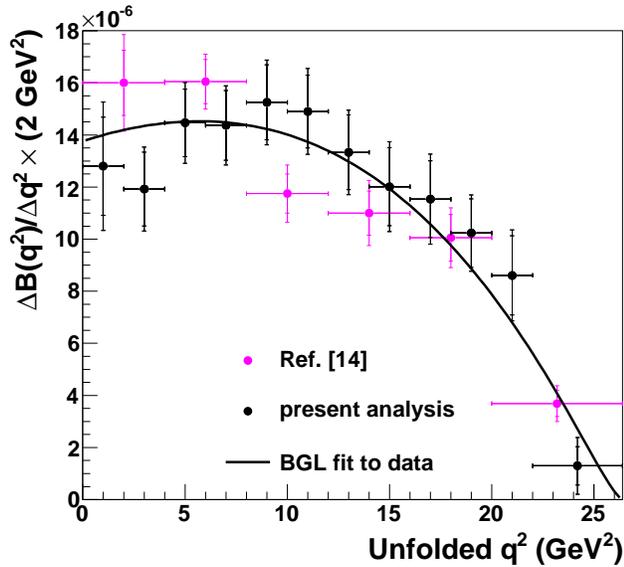,height=8.1cm}
\caption[]{\label{figCombined} 
(color online) Partial \bfpilnuqq\ spectrum for \pilnu\ decays, in 12 bins of 
\qq\ for the present work and six bins of \qq\ for Ref.~\cite{Jochen}. The 
smaller error bars are statistical only while the larger ones also include 
systematic uncertainties. The solid black curve shows the result of the fit to 
the combined data for the two analyses using the BGL~\cite{BGL} 
parametrization. }
\end{center}
\end{figure}

 The above averaging method is not appropriate for the fitted BGL coefficients 
($a_1/a_0$ and $a_2/a_0$) and the value of $|V_{ub}f_{+}(0)|$, since, as shown
in Table~\ref{average}, the two measurements of these quantities are only 
marginally compatible. Instead, we perform a new fit of the BGL parametrization
to the combined partial branching fraction results from the two analyses, the 
12 values obtained in this analysis and the six values from Ref.~\cite{Jochen}.
Here again, the statistical covariance matrices are uncorrelated and the 
systematic covariance matrices are fully correlated between the two data sets. 
The combined error matrix from the two analyses is used to perform the fit, 
with the result shown in Fig.~\ref{figCombined} and a $\chi^2$ probability 
$P(\chi^2)=14.2\%$. When only the statistical covariance matrix is used, the 
$\chi^2$ probability is reduced to 3.1\%. We note that the discrepancy in the 
two analyses of the partial BFs at low values of \qq\ does not lead to 
discrepancies in the resulting values of the total BF or $|V_{ub}|$, as is 
evident in Table~\ref{average}. Finally, we do not attempt to average the 
partial branching fractions due to the different \qq\ binning used in the two 
analyses.   

\section{Summary}

 In summary, we have measured the partial BFs of \etalnu\ decays in three bins 
of \qq\ and of \pilnu\ decays in \nQ\ bins of \qq. From these distributions, we
extract the \fplus\ shapes which are found to be compatible with all three 
theoretical predictions considered for the \pilnu\ decays and with the LCSR 
calculation for the \etalnu\ decays. The BGL parametrization fits our data 
well and allows us to obtain the value of $|V_{ub}f_+(0)|$. Our measured 
branching fractions of the three decays reported in this work lead to a 
significant improvement in our knowledge of the composition of the inclusive 
charmless semileptonic decay rate. Our value of the total BF for \etaplnu\ is 
an order of magnitude smaller than the most recent CLEO 
result~\cite{CLEOpilnu2}. Our value of the total BF for \etalnu\ is consistent 
with a previous untagged \babar\ result~\cite{oldeta}. The value of the ratio 
\bfetaplnu/\bfetalnu\ = $0.67 \pm0.24_{stat} \pm0.11_{syst}$ allows an 
important gluonic singlet contribution to the $\eta^{\prime}$ form factor. The 
present value of the total BF for \pilnu\ is in good agreement with a previous 
untagged \babar\ measurement~\cite{PreviousVub} as well as with a recent 
\babar\ result~\cite{Jochen}. It has comparable precision to the present
world average~\cite{PDG10}. For \pilnu\ decays, we obtain values of~\vub\ for 
three different QCD calculations. The results are in good agreement with those 
of Refs.~\cite{PreviousVub, Jochen}. The three values are all acceptable 
according to the data. Two of these values~\cite{HPQCD06, LCSR} are consistent,
within large theoretical uncertainties, with the value measured in inclusive 
semileptonic $B$ decays: \vub\ = $(4.27 \pm 0.38)\times{10^{-3}}$~\cite{PDG10}.
We also provide the average values of the total BF and of $|V_{ub}|$ obtained 
in the present work and those of Ref.~\cite{Jochen}. We also give the values of
$|V_{ub}f_{+}(0)|$, $a_1/a_0$ and $a_2/a_0$ obtained in a combined BGL fit to 
the two data sets. It may be noted that our results for the \pilnu\ decays are
generally in good agreement with those obtained recently by the Belle 
collaboration~\cite{Belle}.  

\newpage
 We would like to thank A. Khodjamirian, A. Kronfeld and R. Van de Water 
for useful discussions concerning their form-factor calculations and for 
providing the values of $\Delta \zeta$ used in this work. We would also
like to thank J. Dingfelder and B. Viaud for their numerous contributions to 
this analysis. We are grateful for the extraordinary contributions of our 
\pep2\ colleagues in achieving the excellent luminosity and machine conditions 
that have made this work possible. The success of this project also relies 
critically on the 
expertise and dedication of the computing organizations that support \babar.
The collaborating institutions wish to thank SLAC for its support and the kind 
hospitality extended to them. This work is supported by the US Department of 
Energy and National Science Foundation, the Natural Sciences and Engineering 
Research Council (Canada), the Commissariat \`a l'Energie Atomique and Institut
National de Physique Nucl\'eaire et de Physique des Particules (France), the
Bundesministerium f\"ur Bildung und Forschung and Deutsche 
Forschungsgemeinschaft (Germany), the Istituto Nazionale di Fisica Nucleare 
(Italy), the Foundation for Fundamental Research on Matter (The Netherlands),
the Research Council of Norway, the Ministry of Education and Science of the 
Russian Federation, Ministerio de Ciencia e Innovaci\'on (Spain), and the
Science and Technology Facilities Council (United Kingdom). Individuals have 
received support from the Marie-Curie IEF program (European Union), the A. P. 
Sloan Foundation (USA) and the Binational Science Foundation (USA-Israel).

\vspace{1cm}
\section{Appendix}

  In Tables~\ref{cutSummaryPi}, \ref{cutSummaryEta} and \ref{cutSummary}, we 
give the functions describing the \qq\ dependence of the selections used to 
reduce the backgrounds in the three decays under study. 

 The list of all the systematic uncertainties, as well as their values for the 
partial and total BFs, are given in Tables~\ref{pierror} and~\ref{etaerror} for
the \pilnu\ and \etaetaplnu\ decays, respectively. In Table~\ref{pierror}, we 
have one column for each bin of \qqr\ and three columns for various ranges of 
\qqr\ as well as the last column for the global result. In row 1, ``Fitted 
yield'', we give the raw fitted yield as the number of events. 
In row 2, ``Yield statistical error'', we give the statistical uncertainty in 
\% for each fitted yield. In row 3, ``Efficiency'', we give the efficiency in 
\% attached to each yield. In row 4, ``Eff. (without FSR)'', we give the 
efficiency in \%, modified to remove the FSR effect. In row 5, ``Unfolded 
yield'', we give the yields from row 1 unfolded to give the true values of the 
yields in each bin, expressed as the number of events. In row 6, ``\DBR", we 
give the values of the partial BFs computed as usual using the true (unfolded) 
yields and the efficiencies with FSR. In row 7, ``\DBR\ (without FSR)", we give
the values of the partial BFs computed as usual using the true (unfolded) 
yields and the efficiencies modified to remove the FSR effect. In rows 8 - 39, 
we give the contributions in \% to the relative systematic uncertainties for 
each value of \DBR\ as a function of \qqr. In row 40, ``Signal MC statistical 
error'', we give the statistical uncertainty due to the number of MC signal 
events. In row 41, ``Total systematic error'', we give the total systematic 
uncertainty in \% for each value of \DBR, obtained as the sum in quadrature of 
all the systematic uncertainties in each column. In row 42, ``Total statistical
error'', we give the statistical uncertainty in \% for each value of \DBR\ 
obtained from propagating the statistical uncertainties on the raw fitted 
yields, following the unfolding process and taking into account the 
efficiencies. In row 43, ``Total error'', we first give the total uncertainty 
in \% for each value of \DBR, obtained as the sum in quadrature of the total 
systematic error and the total statistical error. We then give, in the last 
four columns, the total uncertainties in \% for each range of \qqr, obtained as
the sum in quadrature of the total errors for the appropriate number of \qqr\ 
bins. A similar description applies to Table~\ref{etaerror}. 

\begin{sidewaysfigure*}
\begin{center}
\epsfig{file=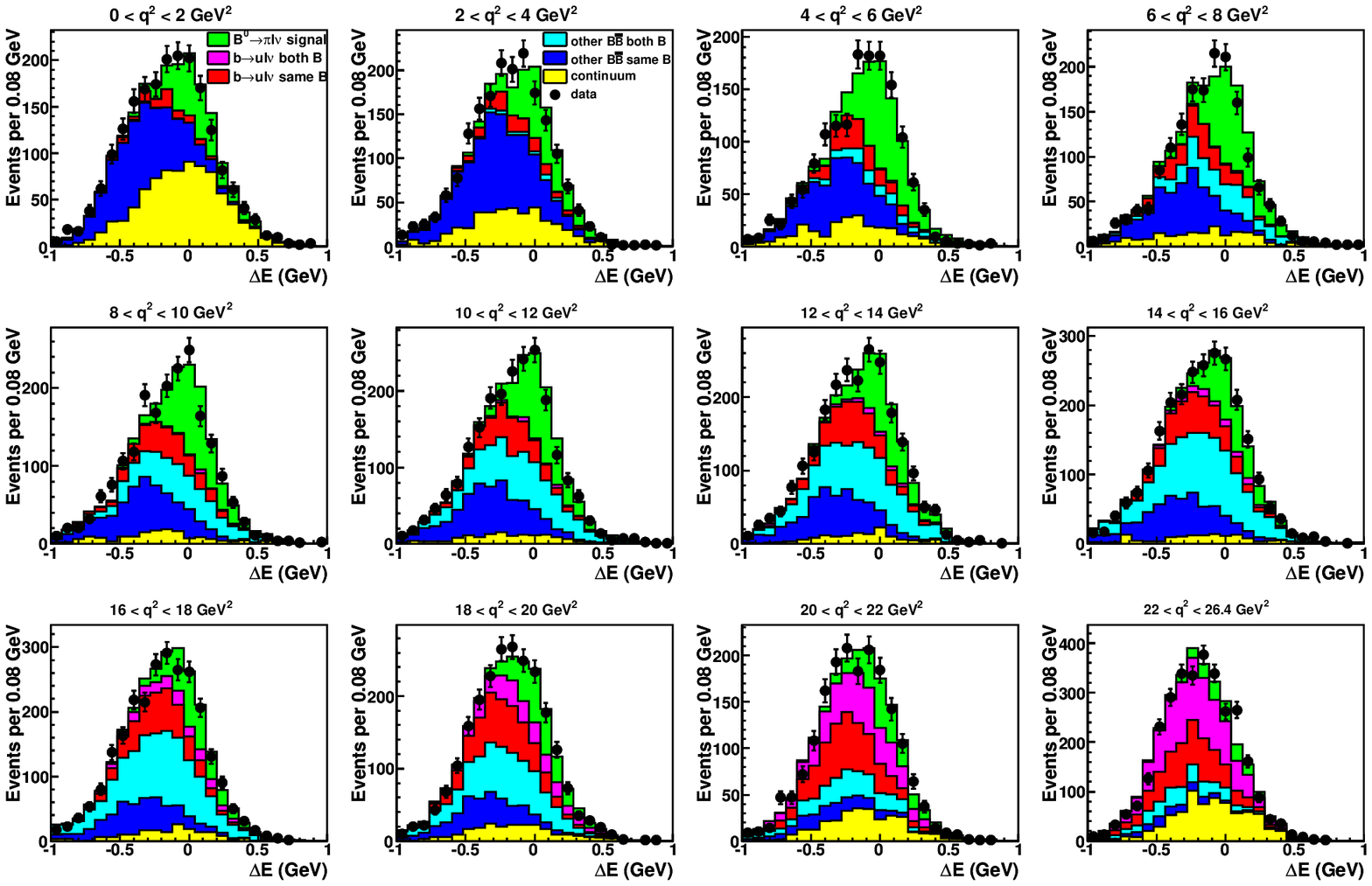,height=15cm}
\caption[]
{\label{dEFitProjDataPi} 
(color online) \DeltaE\ yield fit projections in the signal-enhanced region, 
with \mes\ $>$ 5.2675 \gev, obtained in 12~\qq~bins from the 
fit to the experimental data for \pilnu\ decays. The fit was done using the 
full \DeltaE-\mes fit region.}
\end{center}
\end{sidewaysfigure*}

\begin{sidewaysfigure*}
\begin{center}
\epsfig{file=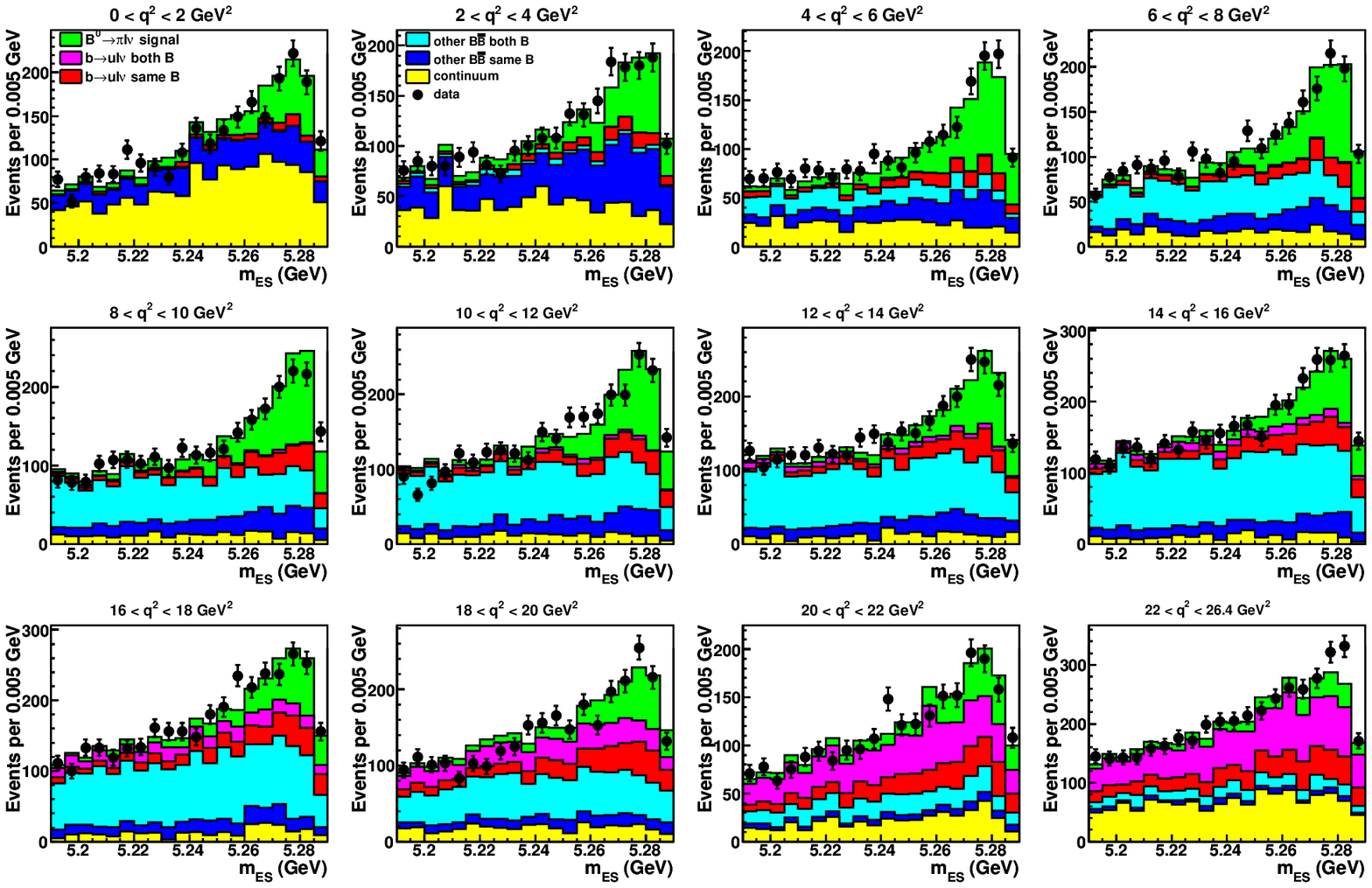,height=15cm}
\caption[]
{\label{mESFitProjDataPi} 
(color online) \mes\ yield fit projections in the signal-enhanced region, with 
$-0.16<\Delta E<0.2~\gev$, obtained in 12~\qq~bins from the fit to the 
experimental data for \pilnu\ decays. The fit was done using the full 
\DeltaE-\mes\ fit region.}
\end{center}
\end{sidewaysfigure*}

 In our analysis, we compute the covariance matrix for each source of 
uncertainty, and use these matrices to calculate the uncertainties on the total
BFs. The correlation matrices for the total statistical and systematic 
uncertainties are given in Table~\ref{StatCovMCEta} for the \etalnu\ yields and
in Tables~\ref{StatCovMC} and~\ref{SystCovMC} for the \pilnu\ yields, 
respectively. Finally, detailed \DeltaE\ and \mes\ fit projections in each 
\qqr\ bin are also shown in Figs.~\ref{dEFitProjDataPi} 
and~\ref{mESFitProjDataPi}, respectively, for the \pilnu\ decays.

\begin{table*}
\caption[]{
\label{cutSummaryPi} \qq-dependent selections used in \pilnu\ decays.}
\begin{center} 
\begin{tabular}{p{10cm}}
\hline\hline
$\cos\theta_{\ell} < 0.85$ for all values of \qqr\ \\
$\cos\theta_{\ell} >-0.0000167*q^8+0.000462*q^6+0.000656*q^4-0.0701*q^2-0.48$\\
$m^2_{miss}/2E_{miss} > -0.5$~\gev\ for all values of \qqr\ \\
$m^2_{miss}/2E_{miss} < 0.00544*q^4-0.127*q^2+1.37$~\gev \\
$\cos\theta_{thrust} < 0.9$ for all values of \qqr\ \\
$\cos\theta_{thrust} < -0.00159*q^4+0.0451*q^2+0.59$ \\
$\theta_{miss} > -0.000122*q^6+0.00483*q^4-0.0446*q^2+0.405$ rad \\
(\qqr\ is given in units of \gevsq) \\
\hline\hline
\end{tabular}
\end{center}
\end{table*}

\begin{table*}
\caption[]{
\label{cutSummaryEta} \qq-dependent selections used in \etalnu\ decays.}
\begin{center}
\begin{tabular}{p{10cm}}
\hline\hline
$\cos\theta_{\ell} < 0.9$ for all values of \qqr\ \\
$\cos\theta_{\ell} > 0.00629*q^4-0.119*q^2-0.252$ \\
$m^2_{miss}/2E_{miss} < 0.8$~\gev, \qqr\ $<$ 7.5 \gevsq\ \\
$m^2_{miss}/2E_{miss} < -0.05*q^2+1.175$~\gev, 7.5 $<$ \qqr\ $<$16.0 \gevsq\ \\
$\cos\theta_{thrust} < 0.05*q^2+0.6$, \qqr\ $<$ 5.0 \gevsq\ \\
$\cos\theta_{thrust} < 0.85$, 5.0 $<$ \qqr\ $<$ 16.0 \gevsq\ \\
$\cos\theta_{miss} < 0.92$, \qqr\ $<$ 11.0 \gevsq\ \\
$\cos\theta_{miss} < 0.88$, 11.0 $<$ \qqr\ $<$ 16.0 \gevsq\ \\
(\qqr\ is given in units of \gevsq) \\
\hline\hline
\end{tabular}
\end{center}
\end{table*}

\begin{table*}
\caption[]{
\label{cutSummary} \qq-dependent selections used in \etaplnu\ decays.}
\begin{center}
\begin{tabular}{p{10cm}}
\hline\hline
$m^2_{miss}/2E_{miss} > -0.3$~\gev\ for all values of \qqr\ \\
$m^2_{miss}/2E_{miss} < 0.35*q^2+0.325$~\gev, \qqr\ $<$ 2.5 \gevsq\ \\
$m^2_{miss}/2E_{miss} < 1.2$~\gev, 2.5 $<$ \qqr\ $<$ 4.5 \gevsq \\
$m^2_{miss}/2E_{miss} < -0.1*q^2+1.65$~\gev, \qqr\ $>$ 4.5 \gevsq\ \\
$\cos\theta_{thrust} < 0.05*q^2+0.575$, \qqr\ $<$ 6.5 \gevsq\ \\
$\cos\theta_{thrust} < 0.9$, 6.5 $<$ \qqr\ $<$ 12.5 \gevsq\ \\
$\cos\theta_{thrust} < -0.05*q^2+1.525$, \qqr\ $>$ 12.5 \gevsq\ \\
$\theta_{miss} > -0.1*q^2+0.45$ \rad, \qqr\ $<$ 2.5 \gevsq\ \\
$\theta_{miss} > 0.2$ \rad, 2.5 $<$ \qqr\ $<$ 5.5 \gevsq\ \\
$\theta_{miss} > 0.05*q^2-0.075$ \rad, \qqr\ $>$ 5.5 \gevsq\ \\
(\qqr\ is given in units of \gevsq) \\
\hline\hline
\end{tabular}
\end{center}
\end{table*}

\begin{table*}
\caption[]{\label{pierror} \pilnu\ yields, efficiencies(\%), \DBR\ $(10^{-7})$ 
and their relative uncertainties (\%). The \DBR\ and efficiency values labeled 
``without FSR'' are modified to remove FSR effects. This procedure has no 
sigificant impact on the \DBR\ values.}
\begin{smaller}
\begin{center}
\begin{tabular}{lcccccccccccccccc}
\hline\hline
\qq\ bins (\gevsq) & 0-2 & 2-4 & 4-6 & 6-8 & 8-10 & 10-12 & 12-14 & 14-16 & 16-18& 18-20& 20-22& 22-26.4& \qq$<$12 & \qq$<$16 & \qq$>$16 & 
Total \\ 
\hline
Fitted yield & 894.7& 987.8& 1177.1& 1181.3& 1178.6& 1122.1& 996.1& 884.5& 904.3& 847.5& 729.9& 873.9& 6541.6& 8422.1& 3355.4& 11777.6\\
Yield statistical error & 12.8& 8.1& 6.0& 6.4& 6.7& 7.0& 8.2& 9.8& 10.3& 10.5& 14.0& 21.0& 3.2& 3.6& 7.9& 3.7 \\ 
\hline
Efficiency         &8.34& 9.10& 9.22& 9.09& 8.59& 8.46& 8.53& 8.50& 9.40& 10.52& 11.61& 14.59& -& -& -& -\\
Eff. (without FSR) &8.00& 8.97& 9.15& 9.18& 8.63& 8.53& 8.58& 8.61& 9.45& 10.66& 11.71& 14.70& -& -& -& -\\
\hline
Unfolded yield          & 919.9& 960.7& 1189.6& 1184.5& 1182.9& 1141.5& 1027.3& 929.2& 979.5& 979.9& 905.8& 376.7& 6579.1& 8535.7& 3241.9& 
11777.6  \\
\DBR\ & 122.7& 117.6& 143.6& 145.0& 153.4& 150.2& 134.1& 121.7& 116.0& 103.7& 86.8& 28.7& 832.5& 1088.3& 335.3& 1423.5\\
\DBR\ (without FSR) & 128.0& 119.2& 144.6& 143.7& 152.5& 149.0& 133.3& 120.1& 115.3& 102.3& 86.1& 28.5& 837.1& 1090.5& 332.3& 1422.8\\
\hline
Tracking efficiency   & 3.2& 1.9& 3.1& 2.2& 2.3& 3.9& 2.6& 4.0& 3.5& 1.3& 4.1& 9.4& 2.3& 2.5& 2.9& 2.6 \\
Photon efficiency     & 6.0& 3.4& 2.6& 1.3& 2.2& 2.5& 3.1& 3.0& 5.0& 1.4& 5.1& 24.2& 1.9& 2.2& 4.6& 2.7 \\
\KL\ efficiency       & 0.9& 0.3& 0.6& 0.3& 0.5& 0.4& 0.5& 0.8& 0.6& 0.4& 1.7& 6.8& 0.3& 0.3& 1.0& 0.4 \\
\KL\ production spectrum  & 1.0& 0.6& 1.0& 0.6& 1.1& 1.0& 0.6& 2.7& 1.7& 1.0& 2.0& 8.3& 0.7& 0.9& 1.9& 1.1 \\
\KL\ energy           & 1.0& 0.6& 0.3& 0.2& 0.2& 0.3& 0.2& 0.4& 0.6& 0.7& 0.8& 7.1& 0.2& 0.3& 0.8& 0.3 \\
$\ell$ identification & 4.0& 1.0& 1.2& 1.3& 0.6& 0.6& 1.6& 1.0& 0.9& 1.6& 0.7& 4.9& 0.3& 0.5& 1.1& 0.6 \\
$\pi$ identification  & 0.5& 0.2& 0.2& 0.2& 0.2& 0.2& 0.2& 0.3& 0.3& 0.3& 0.3& 5.5& 0.2& 0.2& 0.7& 0.3 \\
Bremsstrahlung        & 0.6& 0.3& 0.1& 0.2& 0.2& 0.2& 0.2& 0.3& 0.3& 0.3& 0.3& 5.3& 0.2& 0.2& 0.7& 0.3 \\
\hline
\qqr\ continuum shape & 7.9& 1.6& 0.9& 0.3& 0.7& 0.3& 1.1& 0.3& 0.7& 1.1& 1.2& 5.4& 0.9& 0.8& 1.0& 0.7 \\
\hline
\bfpiZlnu\            & 0.5& 0.1& 0.1& 0.1& 0.1& 0.1& 0.2& 0.3& 0.3& 0.3& 0.4& 7.1& 0.2& 0.2& 0.8& 0.3 \\
\bfrhoClnu\           & 0.5& 0.3& 0.1& 0.1& 0.2& 0.1& 0.2& 0.3& 0.3& 0.3& 0.5& 10.1& 0.2& 0.2& 0.9& 0.3 \\
\bfrhoZlnu\           & 0.7& 0.2& 0.1& 0.1& 0.2& 0.2& 0.2& 0.3& 0.4& 0.3& 0.3& 7.3& 0.2& 0.2& 0.8& 0.3 \\
\bfomegalnu\          & 0.5& 0.1& 0.1& 0.1& 0.1& 0.1& 0.2& 0.3& 0.3& 0.4& 0.3& 8.2& 0.2& 0.2& 0.9& 0.3 \\
\bfetalnu\            & 0.5& 0.1& 0.1& 0.1& 0.1& 0.1& 0.2& 0.3& 0.3& 0.3& 0.2& 5.5& 0.2& 0.2& 0.7& 0.3 \\
\bfetaplnu\           & 0.5& 0.1& 0.1& 0.1& 0.1& 0.1& 0.2& 0.3& 0.3& 0.3& 0.3& 5.4& 0.2& 0.2& 0.7& 0.3 \\
Nonresonant \ulnu\ BF& 0.6& 0.3& 0.1& 0.1& 0.2& 0.2& 0.2& 0.4& 0.5& 0.3& 0.6& 7.8& 0.2& 0.2& 0.7& 0.3 \\
SF parameters         & 0.9& 0.5& 0.9& 0.4& 0.3& 0.5& 0.6& 0.3& 0.5& 2.3& 4.1& 23.4& 0.6& 0.4& 2.3& 0.8 \\
\rholnu\ FF           & 2.4& 1.4& 2.1& 1.4& 1.9& 1.6& 0.8& 0.7& 2.4& 3.0& 1.1& 16.6& 1.8& 1.5& 1.8& 1.5 \\
\pilnu\ FF            & 0.5& 0.1& 0.1& 0.1& 0.1& 0.1& 0.2& 0.3& 0.3& 0.3& 0.3& 7.5& 0.2& 0.2& 0.9& 0.3 \\
Other scalar FF       & 1.1& 0.3& 0.5& 0.5& 0.4& 0.3& 0.3& 0.7& 1.7& 2.1& 2.0& 8.7& 0.4& 0.3& 0.6& 0.3 \\
\omegalnu\ FF         & 0.6& 0.1& 0.1& 0.1& 0.1& 0.1& 0.2& 0.3& 0.5& 0.6& 0.6& 18.1& 0.2& 0.2& 1.8& 0.5 \\
\hline
\bfDlnu\              & 0.6& 0.6& 0.2& 0.1& 0.3& 0.2& 0.5& 0.3& 0.4& 0.4& 0.4& 5.6& 0.2& 0.3& 0.7& 0.4 \\
\bfDstrlnu\           & 0.7& 0.3& 0.2& 0.3& 0.4& 0.3& 0.3& 0.5& 0.4& 0.3& 0.5& 5.7& 0.3& 0.3& 0.7& 0.4 \\
\bfDdblstrlnu\        & 0.7& 0.3& 0.3& 0.5& 0.5& 0.3& 0.8& 0.6& 1.1& 0.7& 0.7& 5.8& 0.3& 0.3& 0.9& 0.4 \\
Nonresonant \clnu\ BF& 0.6& 0.2& 0.2& 0.1& 0.2& 0.2& 0.2& 0.5& 0.3& 0.3& 0.3& 5.6& 0.2& 0.2& 0.7& 0.3 \\
\dlnu\ FF             & 0.5& 0.2& 0.3& 0.1& 0.2& 0.1& 0.2& 0.4& 0.4& 0.3& 0.4& 5.7& 0.2& 0.2& 0.7& 0.3 \\
\dstrlnu\ FF          & 0.6& 0.2& 0.2& 0.4& 0.2& 1.0& 0.6& 1.9& 0.4& 0.7& 1.1& 6.9& 0.2& 0.3& 1.0& 0.5 \\
\hline
$\Upsilon(4S)\rightarrow B^0\bar{B^0}$ BF& 1.5& 1.7& 1.2& 1.3& 1.3& 1.2& 1.5& 1.2& 1.4& 1.4& 1.0& 5.8& 1.3& 1.4& 1.2& 1.3 \\
Secondary lepton      & 4.3& 3.2& 2.1& 1.2& 1.7& 0.5& 1.2& 0.5& 0.5& 0.9& 3.7& 5.8& 1.0& 0.9& 1.2& 0.9 \\
Final state radiation & 0.3& 1.3& 0.8& 2.2& 0.3& 1.4& 1.2& 1.3& 1.4& 1.6& 0.8& 3.4& 1.0& 1.1& 1.5& 1.2 \\
B counting            & 1.1& 1.1& 1.1& 1.1& 1.1& 1.1& 1.1& 1.1& 1.1& 1.1& 1.1& 1.1& 1.1& 1.1& 1.1& 1.1 \\
Fit bias              & 0.1& 0.3& 0.4& 0.1& 0.1& 0.4& 0.4& 0.7& 0.1& 1.0& 2.0& 30.8& 0.2& 0.0& 1.8& 0.4 \\
Signal MC stat error  & 1.3& 1.6& 1.4& 1.6& 1.4& 1.5& 1.4& 1.4& 1.3& 1.4& 1.2& 2.5& 0.6& 0.4& 0.6& 0.3 \\
\hline
Total systematic error  & 12.9& 6.4& 6.0& 4.9& 5.0& 5.9& 5.6& 7.0& 7.8& 6.3& 10.0& 61.6& 4.4& 4.6& 8.3& 5.2 \\ \hline
Total statistical error & 14.7& 11.9& 9.0& 9.3& 9.4& 9.4& 10.8& 12.5& 12.8& 12.8& 17.6& 56.7& 3.9& 3.7& 7.6& 3.5 \\ \hline
Total error & 19.6& 13.5& 10.8& 10.5& 10.7& 11.1& 12.1& 14.3& 15.0& 14.3& 20.3& 83.8& 5.9& 5.9& 11.3& 6.3\\
\hline\hline
\end{tabular}
\end{center}
\end{smaller}
\end{table*}

\begin{table*}
\caption[]{\label{etaerror} \etaetaplnu\ yields, efficiencies(\%), \DBR\ $(10^{-7})$ and their relative uncertainties (\%).}
\begin{smaller}
\begin{center}
\begin{tabular}{lcp{0.2cm}cp{0.2cm}ccccp{0.3cm}cccc}
\hline\hline
Decay mode & $\eta^{\prime}\ell^+\nu$ & & $\eta\ell^+\nu$ ($3\pi$)& & \multicolumn{4}{c}{$\eta\ell^+\nu$ ($\gamma\gamma$)}& & \multicolumn{4}{c}{$\eta\ell^+\nu$ ($3\pi$ and $\gamma\gamma$ combined)}\\
\qq\ bins (\gevsq)      & Total& &Total& &0-4  & 4-8  & 8-16 & Total& &0-4  & 4-8  & 8-16 & Total \\ 
\hline
Fitted yield           & 141.0& &244.8& &279.9& 216.8& 146.7& 643.4& &303.9& 331.5& 252.5& 887.9\\
Yield statistical error              & 32.8 & &25.6 & &13.9 & 17.2 & 33.9 & 12.0 & &14.1 & 14.2 & 26.6 & 11.0\\
\hline
Efficiency             & 0.61 & &0.59 & &2.01 & 2.55 & 1.42 & -    & &2.53 & 3.41 & 1.94 & - \\
\hline
Unfolded yield         & 141.0& &244.8& &299.1& 210.9& 133.3& 643.4& &319.3& 334.8& 233.9& 887.9\\
\DBR\                  & 242.5& &431.5& &155.3& 86.3 & 97.7 & 339.3& &131.8& 102.6& 126.2& 360.6  \\
\hline
Tracking efficiency    & 5.2  & &4.1  & &3.2  & 2.4  & 14.6 & 2.6  & &2.1  & 2.0  & 11.1 & 2.8 \\
Photon efficiency      & 5.6  & &3.1  & &10.1  & 4.3  & 27.4 & 7.0  & &8.0  & 3.8  & 9.0  & 5.7 \\
\KL\ efficiency        & 2.5  & &0.7  & &8.6  & 2.9  & 27.2  & 3.2  & &1.0  & 0.5  & 2.2  & 0.6 \\
\KL\ production spectrum & 2.7  & &1.4  & &4.7  & 1.5  & 16.2  & 2.5  & &0.8  & 0.5  & 2.3  & 1.0 \\
\KL\ energy            & 1.1  & &1.4  & &0.6  & 0.5  & 2.5  & 0.9  & &0.6  & 0.4  & 2.3  & 1.0 \\
$\ell$ identification  & 2.0  & &1.8  & &0.1  & 2.7  & 3.9  & 1.8  & &0.2  & 1.9  & 3.4  & 1.8 \\
$\pi$ identification   & 0.6  & &0.5  & &-    & -    & -    & -    & &0.1  & 0.2  & 0.5  & 0.3 \\
Bremsstrahlung         & 0.5  & &0.2  & &1.6  & 2.7  & 22.2 & 8.0  & &0.3  & 0.7  & 12.3 & 4.2 \\
\hline
Continuum yield        & 4.9  & &1.1  & &-    & -    & -    & -    & &-    & -    & -    & -   \\
\qqr\ continuum shape  & 5.2  & &2.6  & &2.6  & 1.5  & 4.5  & 0.5  & &2.4  & 0.7  & 2.8  & 0.3 \\
\hline
\bfpilnu\              & 0.0  & &0.1  & &0.0  & 0.0  & 0.1  & 0.0  & &0.0  & 0.0  & 0.1  & 0.0 \\
\bfpiZlnu\             & 0.2  & &0.0  & &0.4  & 0.9  & 5.2  & 1.9  & &0.3  & 0.6  & 2.9  & 1.3 \\
\bfetaetaplnu\         & 0.4  & &0.4  & &0.0  & 0.1  & 0.8  & 0.2  & &0.1  & 0.1  & 1.0  & 0.4 \\
\bfrhoClnu\            & 0.3  & &0.5  & &0.1  & 1.1  & 6.9  & 2.3  & &0.1  & 0.6  & 4.2  & 1.7 \\
\bfrhoZlnu\            & 0.0  & &0.3  & &0.1  & 0.1  & 0.5  & 0.1  & &0.0  & 0.1  & 0.8  & 0.2 \\
\bfomegalnu\           & 0.8  & &1.1  & &0.1  & 0.2  & 2.6  & 0.8  & &0.1  & 0.1  & 2.6  & 0.9 \\
Nonresonant \ulnu\ BF & 2.3  & &3.5  & &0.4  & 0.9  & 9.5  & 3.1  & &0.5  & 0.6  & 8.6  & 3.4 \\
$\eta$ BF              & 3.1  & &1.2  & &0.5  & 0.7  & 0.7  & 0.6  & &0.5  & 0.6  & 0.7  & 0.5 \\
SF parameters          & 4.3  & &6.3  & &1.4  & 2.7  & 16.8 & 6.1  & &1.5  & 2.5  & 14.3 & 6.2 \\
\rholnu\ FF            & 0.1  & &0.7  & &0.1  & 2.3  & 1.7  & 0.9  & &0.1  & 1.5  & 0.9  & 0.5 \\
\etaetaplnu\ FF        & 1.1  & &1.0  & &0.1  & 0.1  & 1.4  & 0.4  & &0.1  & 0.1  & 1.5  & 0.6 \\
Other scalar FF        & 2.9  & &4.2  & &7.7  & 1.4  & 0.1  & 3.2  & &0.7  & 0.1  & 0.0  & 0.2 \\
\omegalnu\ FF          & 1.2  & &2.1  & &0.1  & 0.5  & 2.8  & 0.7  & &0.1  & 0.4  & 3.9  & 1.3 \\
\hline
\bfDlnu\               & 1.6  & &0.7  & &0.3  & 0.7  & 0.6  & 0.3  & &0.3  & 0.7  & 0.7  & 0.4 \\
\bfDstrlnu\            & 0.3  & &0.4  & &0.1  & 0.8  & 1.2  & 0.4  & &0.1  & 0.7  & 1.0  & 0.4 \\
\bfDdblstrlnu\         & 2.0  & &1.2  & &0.6  & 0.9  & 2.5  & 0.7  & &0.6  & 0.7  & 2.6  & 0.9 \\
Nonresonant \clnu\ BF & 0.1  & &0.1  & &0.2  & 0.1  & 0.8  & 0.2  & &0.3  & 0.1  & 0.4  & 0.2 \\
\dlnu\ FF              & 0.1  & &0.3  & &0.1  & 0.1  & 0.5  & 0.2  & &0.1  & 0.1  & 0.7  & 0.3 \\
\dstrlnu\ FF           & 0.6  & &0.9  & &0.5  & 0.9  & 1.3  & 0.4  & &0.5  & 1.2  & 1.2  & 0.4 \\
\hline
${\BR}(\Upsilon(4S)\rightarrow B^0\bar{B^0})$ 
                       & 1.1  & &1.2  & &1.4  & 1.1  & 0.9  & 1.2  & &1.4  & 1.2  & 1.0  & 1.2 \\
Secondary lepton       & 4.2  & &5.0  & &1.3  & 0.7  & 9.1  & 2.1  & &1.2  & 1.6  & 9.3  & 3.0 \\
B counting             & 1.1  & &1.1  & &1.1  & 1.1  & 1.1  & 1.1  & &1.1  & 1.1  & 1.1  & 1.1\\
Signal MC stat error   & 1.2  & &1.1  & &1.4  & 1.6  & 1.2  & 0.7  & &1.3  & 1.3  & 1.0  & 0.5 \\
\hline 
Total systematic error & 14.3 & &12.4 & &17.0 & 8.7  & 55.4 & 14.1 & &9.3  & 6.6  & 28.7 & 11.6 \\ \hline
Total statistical error & 32.8 & &25.6 & &14.6 & 21.0 & 39.3 & 13.7 & &15.2 & 16.6 & 30.3 & 12.5 \\ \hline
Total error            & 35.8 & &28.4 & &22.4 & 22.7 & 67.9 & 19.6 & &17.8 & 17.8 & 41.8 & 17.0 \\
\hline\hline
\end{tabular}
\end{center}
\end{smaller}
\end{table*}

\begin{table*}
\caption[]{\label{StatCovMCEta} Correlation matrix of the partial \bfetalnuq\ 
statistical and systematic uncertainties.}
\begin{center}
\begin{smaller}
\begin{tabular}{lp{0.1cm}cccp{0.2cm}ccc}
\hline\hline
 & &\multicolumn{3}{c}{Statistical} & & \multicolumn{3}{c}{Systematic}\\
\qq\ bins (\gevsq)& & 0-4 & 4-8 & 8-16 & &0-4 & 4-8 & 8-16 \\ \hline
0-4& & 1.00 & -0.08 & 0.00 & &1.00 & 0.36 & 0.05\\
4-8& & -0.08 & 1.00 & -0.06 & &0.36 & 1.00 & 0.29\\
8-16& & 0.00 & -0.06& 1.00 & &0.05 & 0.29 & 1.00\\ \hline\hline
\end{tabular}
\end{smaller}
\end{center}
\end{table*}

\begin{table*}
\caption[]{\label{StatCovMC} Correlation matrix of the partial \bfpilnuq\ 
statistical uncertainties.}
\begin{center}
\begin{smaller}
\begin{tabular}{lcccccccccccc}
\hline\hline
\qq\ bins (\gevsq) & 0-2 & 2-4 & 4-6 & 6-8 & 8-10 & 10-12 & 12-14 & 14-16 & 16-18& 18-20 & 20-22 & 22-26.4\\ \hline
0-2 & 1.00 & -0.16 & 0.17 & 0.02 & -0.02 & 0.03 & 0.01 & 0.04 & 0.05 & 0.02 & 0.04 & -0.00\\
2-4 & -0.16 & 1.00 & -0.32 & 0.11 & 0.00 & -0.00 & -0.01 & 0.01 & 0.01 & -0.00 & 0.00 & -0.00\\
4-6 & 0.17 & -0.32 & 1.00 & -0.30 & 0.15 & 0.02 & 0.06 & 0.06 & 0.07 & 0.00 & 0.01 & 0.01\\
6-8 & 0.02 & 0.11 & -0.30 & 1.00 & -0.22 & 0.13 & 0.07 & 0.06 & 0.07 & 0.00 & 0.00 & 0.02\\
8-10 & -0.02 & 0.00 & 0.15 & -0.22 & 1.00 & -0.22 & 0.16 & 0.05 & 0.08 & 0.01 & -0.00 & 0.02\\
10-12 & 0.03 & -0.00 & 0.02 & 0.13 & -0.22 & 1.00 & -0.15 & 0.10 & 0.07 & -0.01 & 0.02 & 0.00\\
12-14 & 0.01 & -0.01 & 0.06 & 0.07 & 0.16 & -0.15 & 1.00 & -0.16 & 0.13 & -0.01 & 0.05 & -0.00\\
14-16 & 0.04 & 0.01 & 0.06 & 0.06 & 0.05 & 0.10 & -0.16 & 1.00 & -0.01 & 0.01 & -0.02 & -0.02\\
16-18 & 0.05 & 0.01 & 0.07 & 0.07 & 0.08 & 0.07 & 0.13 & -0.01 & 1.00 & -0.17 & 0.09 & -0.08\\
18-20 & 0.02 & -0.00 & 0.00 & 0.00 & 0.01 & -0.01 & -0.01 & 0.01 & -0.17 & 1.00 & 0.05 & -0.05\\
20-22 & 0.04 & 0.00 & 0.01 & 0.00 & -0.00 & 0.02 & 0.05 & -0.02 & 0.09 & 0.05 & 1.00 & -0.35\\
22-26.4 & -0.00 & -0.00 & 0.01 & 0.02 & 0.02 & 0.00 & -0.00 & -0.02 & -0.08 & -0.05 & -0.35 & 1.00\\
\hline\hline
\end{tabular}
\end{smaller}
\end{center}
\end{table*}

\begin{table*}
\caption[]{\label{SystCovMC} Correlation matrix of the partial \bfpilnuq\ 
systematic uncertainties.}
\begin{center}
\begin{smaller}
\begin{tabular}{lcccccccccccc}
\hline\hline
\qq\ bins (\gevsq) & 0-2 & 2-4 & 4-6 & 6-8 & 8-10 & 10-12 & 12-14 & 14-16 & 16-18& 18-20 & 20-22 & 22-26.4\\ \hline
0-2 & 1.00 & -0.45 & 0.37 & 0.30 & 0.59 & 0.47 & 0.54 & 0.38 & 0.39 & 0.03 & 0.44 & 0.34\\
2-4 & -0.45 & 1.00 & -0.24 & 0.03 & -0.27 & -0.09 & -0.17 & -0.19 & -0.37 & 0.36 & -0.37 & -0.02\\
4-6 & 0.37 & -0.24 & 1.00 & 0.78 & 0.83 & 0.76 & 0.67 & 0.68 & 0.52 & 0.31 & 0.71 & 0.42\\
6-8 & 0.30 & 0.03 & 0.78 & 1.00 & 0.71 & 0.74 & 0.65 & 0.63 & 0.46 & 0.47 & 0.53 & 0.32\\
8-10 & 0.59 & -0.27 & 0.83 & 0.71 & 1.00 & 0.74 & 0.71 & 0.70 & 0.50 & 0.35 & 0.66 & 0.38\\
10-12 & 0.47 & -0.09 & 0.76 & 0.74 & 0.74 & 1.00 & 0.74 & 0.80 & 0.61 & 0.36 & 0.61 & 0.38\\
12-14 & 0.54 & -0.17 & 0.67 & 0.65 & 0.71 & 0.74 & 1.00 & 0.69 & 0.73 & 0.29 & 0.56 & 0.33\\
14-16 & 0.38 & -0.19 & 0.68 & 0.63 & 0.70 & 0.80 & 0.69 & 1.00 & 0.71 & 0.34 & 0.65 & 0.36\\
16-18 & 0.39 & -0.37 & 0.52 & 0.46 & 0.50 & 0.61 & 0.73 & 0.71 & 1.00 & -0.03 & 0.62 & 0.22\\
18-20 & 0.03 & 0.36 & 0.31 & 0.47 & 0.35 & 0.36 & 0.29 & 0.34 & -0.03 & 1.00 & -0.02 & 0.18\\
20-22 & 0.44 & -0.37 & 0.71 & 0.53 & 0.66 & 0.61 & 0.56 & 0.65 & 0.62 & -0.02 & 1.00 & 0.52\\
22-26.4 & 0.34 & -0.02 & 0.42 & 0.32 & 0.38 & 0.38 & 0.33 & 0.36 & 0.22 & 0.18 & 0.52 & 1.00\\
\hline\hline
\end{tabular}
\end{smaller}
\end{center}
\end{table*}

\begin{thebibliography}{99}
\bibitem{PlusCC}
Charge conjugate decays and $\ell = e$ or $\mu$ are implied throughout this 
paper.
\end{thebibliography}
\end{document}